\documentclass{aa}  
\usepackage{graphicx}
\usepackage{txfonts}
\usepackage{lipsum}
\usepackage{lscape} 
\usepackage{xcolor}
\usepackage{placeins}
\usepackage[colorlinks=true,
            linkcolor=blue,
            citecolor=blue,
            urlcolor=blue]{hyperref}
\begin{document}
\nolinenumbers
\title{The Deepest X-ray View of a Kiloparsec Scale Jet at $z=6.1$}
\author{F. A. Meanti\inst{1,2}
        \and L. Ighina\inst{3,1}\fnmsep\thanks{Corresponding author: luca.ighina@cfa.harvard.edu}
        \and A. Moretti\inst{1}
        \and K. Bettwy\inst{3,4}
        \and A. Caccianiga\inst{1}
        \and T. Connor\inst{3}
        \and F. Tavecchio\inst{1}
        \and S. Belladitta\inst{5,6}
        \and E. K. Bempong-Manful\inst{7,8}
        \and C. Spingola\inst{9}
        }
\institute{INAF, Osservatorio Astronomico di Brera, via Brera 28, 20121, Milano, Italy 
\and Politecnico di Milano, Department of Aerospace Science and Technology, Via La Masa 34, 20156, Milano, Italy
\and Center for Astrophysics | Harvard \& Smithsonian, 60 Garden St., Cambridge, MA 02138, USA
\and Department of Astronomy, Yale University, 219 Prospect Street, New Haven, CT 06511, USA
\and Max Planck Institut für Astronomie, Königstuhl 17, D-69117 Heidelberg, Germany
\and INAF – Osservatorio di Astrofisica e Scienza dello Spazio di Bologna, Via Gobetti 93/3, I-40129 Bologna, Italy
\and Jodrell Bank Centre for Astrophysics, Department of Physics and Astronomy, The University of Manchester, Manchester M13 9PL, UK
\and School of Physics, University of Bristol, Tyndall Avenue, Bristol BS8 1TL, UK
\and INAF $-$ Istituto di Radioastronomia, Via Gobetti 101, I$-$40129, Bologna, Italy
}
\date{Received October 13, 2026}
\abstract
{{\it Context:} The mechanism responsible for the high-energy emission observed in kiloparsec scale jets across cosmic time remains poorly understood. The relativistic jet associated with the $z=6.1$ AGN PSO~J030947.49+271757.31 represents the highest-$z$ large-scale jet currently resolved in the X-rays and provides a unique laboratory to study the evolution of these structures. \\
{\it Aim:} We present a study aimed at constraining the morphology and spectral properties the kpc-scale jet associated with PSO~J030947.49+271757.31. In addition, we investigate the capabilities of future X-ray observatories for the study of extended jets.\\
{\it Methods:} We obtained new dedicated observations of PSO~J030947.49+271757.31 with the {\it Chandra X-ray Observatory} (for a total of 360 ksec combined with archival data). Thanks to the improvement of a factor of $\sim3$ in exposure time compared to previous studies, we spatially resolved the extended X-ray emission into two distinct components and performed spatially resolved spectral analyses of the different regions.\\ 
{\it Results:} We confirmed the presence of extended X-ray emission of PSO~J030947.49+271757.31 with significantly improved statistics and we identified two separate and slightly resolved components. The X-ray spectrum of the extended emission has a photon index of $\Gamma = 2.25^{+0.45}_{-0.44}$, steeper than that of the core ($\Gamma = 1.86\pm0.09$), and an extended-to-core luminosity ratio of $L_{\rm ext}/L_{\rm core}=18\%$. Using the {\it Chandra} image as input, we performed simulations for next-generation X-ray facilities. An AXIS-like mission would increase the detected photon statistics by a factor of $\sim17$ compared to the current {\it Chandra} dataset.\\
{\it Conclusions:} We resolved the kpc-scale jet associated with this $z=6.1$ AGN into two main components and constrained its spectral properties. This extended component corresponds to the most distant and one of the most X-ray-luminous large-scale jets currently known. Simulations show that future X-ray missions with arcsecond-scale angular resolution will not only enable significantly more detailed studies of systems like PSO~J030947.49+271757.31, but will also reveal a population of fainter high-redshift jets currently beyond the reach of {\it Chandra}.\\
}

\maketitle
\nolinenumbers
\section{Introduction}

Relativistic jets, generated by accreting supermassive black holes (SMBHs), are an essential part of Active Galactic Nuclei (AGN) evolution and feedback. They can transport a large amount of energy and release it onto the host galaxy and surrounding intergalactic medium \citep[see][]{Fabian2012,Blandford2019}.
Observations with the {\it Chandra X-ray Observatory} showed that most of the extended radio jets from AGN have high-energy counterparts detectable up to hundreds of kpc away from the central SMBH (e.g., \citealt{Schwartz2000,Sambruna2004,Connor2021}). However, despite the detection of hundreds of such systems \citep[e.g.][]{Harris2006}, these data have not yielded a homogeneous consensus in their interpretation but rather has given rise to a controversy with profound implications for relativistic jet composition and physics (e.g., \citealt{Georganopoulos2006,Tavecchio2021,Meyer2023}).

Based on one of the first and most popular scenarios, high-energy radiation observed in kpc-scale jets is produced through the Inverse Compton interaction of the electrons within the jet with the Cosmic Microwave Background (CMB) photons (IC/CMB), which are up-scattered into the X-ray and $\gamma$-ray band (\citealt{Tavecchio2000,Celotti2001}). 
However, the non-detection of a permanent emission by the {\it Fermi}-LAT detector in many sources at low redshift ($z<1$; e.g., \citealt{Breiding2017,Breiding2023}) implies that this process cannot be the main contributor to the observed X-ray emission in local sources, implying the presence of a second, highly energetic population of electrons (e.g, \citealt{Georganopoulos2006, Tavecchio2021,Rahman2024}).

Nonetheless, some level of IC/CMB is necessarily produced, especially at high redshift, due to the strong evolution of the CMB energy density, (1+$z$)$^4$ (e.g., \citealt{Ighina2021b}). Moreover, the $\gamma$-ray levels measured in a few local systems are actually consistent with the expectations from the IC/CMB (e.g. \citealt{Meyer2019}). Interestingly, these sources are blazars, that is, with relativistic jets oriented close to our line of sight. For these reasons, we expect the IC/CMB to dominate the high-energy emission of kpc-scale jets of high-$z$ blazars. From the characterization of these systems we can then extrapolate the expected IC/CMB contribution to lower redshift, $\propto$(1+$z$)$^4$, and therefore put constraints on the properties of the highly energetic population of electrons responsible for the majority of the X-ray emission at lower redshift (e.g. \citealt{Tavecchio2007,Worrall2020,Liniewicz2026}).
At the same time, blazars also represent an important tool for constraining the evolution of the overall jetted AGN population across redshift. By definition blazars are detected at small viewing angles from the jet axis ($\theta_{\rm view}<1/\Gamma$, $\Gamma\sim10$, the bulk Lorentz factor; \citealt{Lister2019}). Therefore, from the detection of a single blazar we can infer the presence of $N_{tot}=2\Gamma^2N_{blazar}$ sources at the same redshift with similar properties but jets oriented in different directions. \citep[e.g.][]{Ghisellini2010,Diana2022}.

In this context, the blazar PSO J030947.49+271757.31 at $z=6.10$ (PSO~J0309+27 hereafter; \citealt{Belladitta2020}) represents one of the best laboratories for investigating the properties and evolution of large-scale relativistic jets. Previous high-resolution very long baseline interferometry (VLBI) radio observations revealed a powerful relativistic jet on $\sim$500~pc scales oriented close to our line of sight \citep{Spingola2020}, while a 128~ksec {\it Chandra} observation identified extended X-ray emission spatially aligned with the radio jet axis, providing the first evidence for a kpc-scale X-ray jet at $z>6$ \citep{Ighina2022a}.

In this work, we present deeper {\it Chandra} X-ray observations of PSO~J0309+27 that allowed for a better morphological and spectral analysis of its extended component.
In Sec. \ref{sec:data} we present the new \textit{Chandra} observations available and describe the imaging reduction process. In Sec. \ref{sec:chandra_spectra} we present the spectral analysis of the {\it Chandra} observations, for both the core and the extended emission. In Sec. \ref{sec:comparison} we compare the X-ray properties of the extended jet associated with PSO~J0309+27 to other kpc-scale jets known in the literature. In Sec. \ref{sec:AXIS} we use simulations to show how a potential next-generation X-ray mission with arcsec angular resolution could significantly improve the study of kpc-scale jets. Finally, in Sect. \ref{sec:conclusions} we summarize our results and conclusions.\\

Throughout the paper we assume a flat $\Lambda$ cold dark matter cosmology with $H_{0}$=70 km s$^{-1}$ Mpc$^{-1}$, $\Omega_m$=0.3, and $\Omega_{\Lambda}$=0.7 \citep[e.g.][]{Spergel2007}, where 1\arcsec\, corresponds to a projected distance of 5.66~kpc at $z$=6.1. Spectral indices are given assuming $S_{\nu}\propto \nu^{-\alpha}$ and the photon index parameter is given by $\Gamma=\alpha+1$. All uncertainties are reported at 90\% confidence unless otherwise specified.

\section{Chandra X-ray observations}
\label{sec:data}
The X-ray data of PSO~J0309+27 consist of 22 observations performed with the \textit{Chandra X-ray Observatory} using the Advanced CCD Imaging Spectrometer (ACIS; \citealt{Garmire2003}) in VFAINT tsmode.
All the observations were performed with the source located at or close to the ACIS-S aim point. The dataset combines 6 archival observations from Cycle 21 (totaling $128\,\mathrm{ksec}$, presented in \citealt{Ighina2022a}) with 16 new proprietary pointings obtained during Cycle 24. The total effective exposure time is $360.2\,\mathrm{ksec}$, providing the deepest X-ray view of this source to date. A log of the 22 observations, including observation identifiers, dates, and exposure times, is provided in Table~\ref{table:1}. 
\begin{table}[ht!]
\caption{Log of the 22 \textit{Chandra} observations of PSO~J0309+27.} 
\label{table:1}
\centering
\begin{tabular}{c c c c}
\hline\hline
Observation ID & Date & Exposure & Cycle \\
 & (yyyy-mm-dd) & (ksec) & \\
\hline
23107 & 2020-03-24 & 27.1 & 21 \\
24513 & 2020-11-03 & 19.9 & 21 \\
24855 & 2020-11-04 & 8.8  & 21 \\
23830 & 2020-11-05 & 22.1 & 21 \\
24856 & 2020-11-06 & 23.7 & 21 \\
24512 & 2020-11-07 & 28.1 & 21 \\
\hline
28151 & 2023-11-14 & 13.1 & 24 \\
29066 & 2023-11-15 & 15.9 & 24 \\
28570 & 2023-11-17 & 10.1 & 24 \\
28576 & 2023-11-18 & 17.6 & 24 \\
29067 & 2023-11-19 & 19.6 & 24 \\
29068 & 2023-11-19 & 15.9 & 24 \\
28574 & 2023-11-21 & 9.4  & 24 \\
29081 & 2023-11-23 & 10.1 & 24 \\
28575 & 2023-11-24 & 15.1 & 24 \\
28571 & 2023-12-06 & 15.1 & 24 \\
29101 & 2023-12-07 & 14.5 & 24 \\
28572 & 2024-11-18 & 15.1 & 24 \\
30627 & 2024-11-19 & 14.6 & 24 \\
30628 & 2024-11-19 & 15.1 & 24 \\
28573 & 2024-11-27 & 23.6 & 24 \\
30640 & 2024-11-30 & 15.1 & 24 \\
\hline\hline
\end{tabular}
\tablefoot{The dataset is available in the \textit{Chandra} Data Collection 587 (\href{https://doi.org/10.25574/cdc.587}{doi:10.25574/cdc.587}).}
\end{table} 
\subsection{Multi-epoch astrometric registration}
Data reduction was carried out using the \textit{Chandra} Interactive Analysis of Observations (\texttt{CIAO}) software package (v4.16; \citealt{Fruscinone2006}) and the Calibration Database (\texttt{CALDB}) version 4.10.3. 

Given the multi-epoch nature of the dataset, a precise astrometric registration of the 22 individual segments was performed to prevent spatial degradation of the source morphology during the stacking process. For each observation, we performed source detection using the \texttt{wavdetect} algorithm in the $0.5$--$7.0\,\mathrm{keV}$ energy range, using wavelet scales of 2, 4, and 8 pixels. Two bright point sources detected in the field, together with the compact core of PSO~J0309+27, were used as astrometric reference points\footnote{The two point sources used as astrometric references are located at 
RA=$47.44155\,\mathrm{deg}$, Dec=$27.25441\,\mathrm{deg}$ and 
RA=$47.49831\,\mathrm{deg}$, Dec=$27.27215\,\mathrm{deg}$ (J2000).}. 

A relative astrometric alignment was then performed by computing the coordinate transformation matrices with the \texttt{wcs\_match} task and updating the aspect solution and event files using \texttt{wcs\_update}. The resulting corrections were dominated by translational shifts ranging from $\sim0.1''$ to $\sim1.6''$, while no significant rotational offsets were detected between the individual observations. These values are consistent with the nominal absolute positional accuracy of \textit{Chandra}, characterized by a $90\%$ uncertainty radius of $1.18''$ and a $99\%$ limit of $1.74''$ for sources observed near the aimpoint\footnote{\url{https://cxc.cfa.harvard.edu/cal/ASPECT/celmon/}}.

The individual event files were then merged using \texttt{merge\_obs} and further refined to correct for residual absolute pointing offsets by cross-matching the X-ray detections with optical counterparts from the Panoramic Survey Telescope and Rapid Response
System survey (Pan-STARRS; \citealt{Chambers2016}). The required correction consists of a small global translational shift of $\Delta RA=+0.289''$ and $\Delta Dec=-0.207''$ with no significant rotation, indicating an excellent initial astrometric solution and the high quality of the individual observations. After this absolute astrometric refinement, the final X-ray position of PSO~J0309+27 is RA=$47.44785\,\mathrm{deg}$ and Dec$=+27.29925\,\mathrm{deg}$, with a residual uncertainty of $\sim0.1''$, dominated by the statistical error on the X-ray centroid.

\subsection{X-ray imaging and surface brightness}
\begin{figure}[t!]
           \centering
           \includegraphics[width=\hsize]{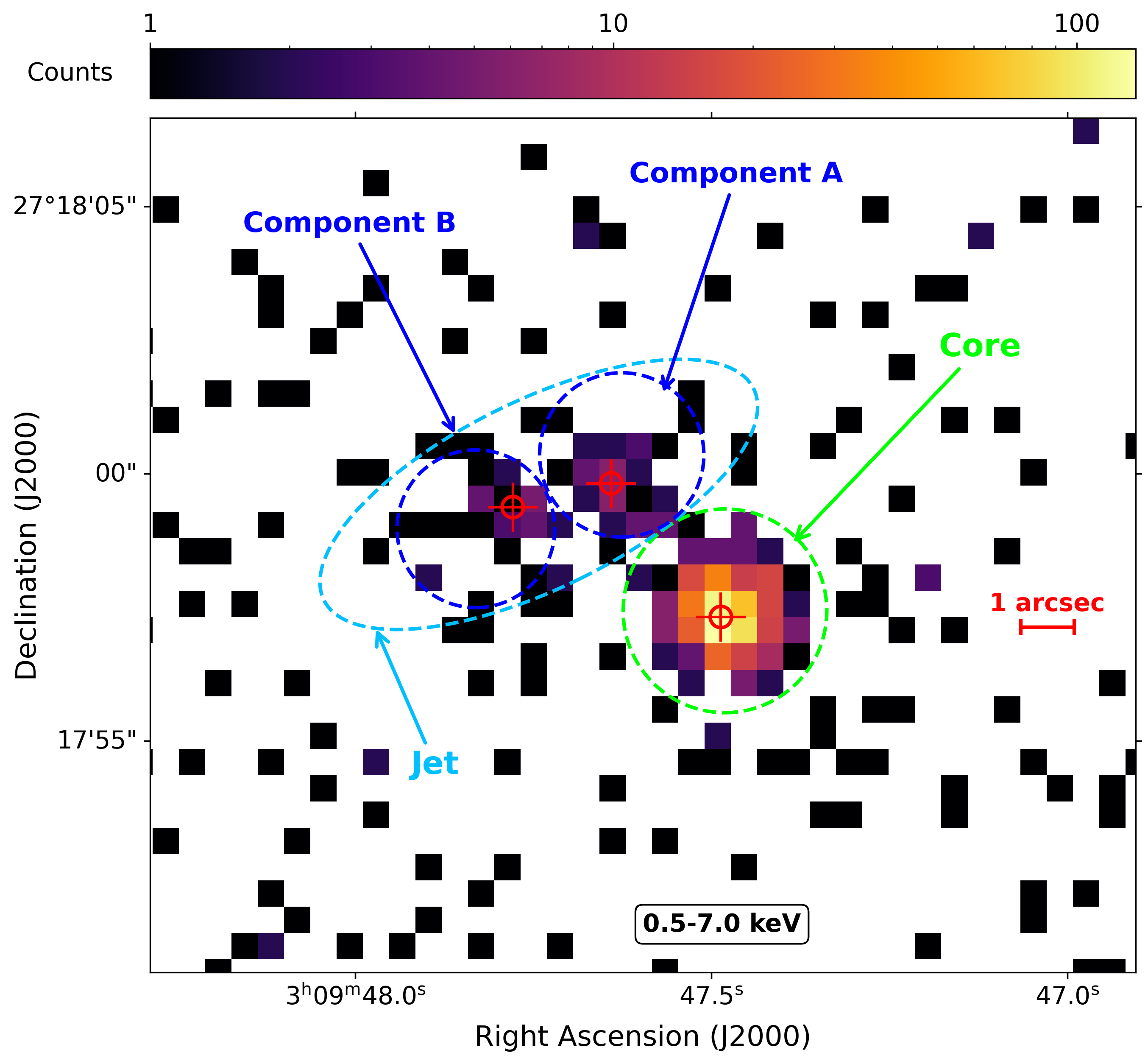}
           \caption{X-ray image of PSO~J0309+27 in the $0.5$--$7.0\,\mathrm{keV}$ energy band obtained from the full \textit{Chandra} dataset (total exposure: $360.2\,\mathrm{ksec}$). The image is displayed using a logarithmic color scale representing counts per pixel. No spatial smoothing or exposure/flux normalization has been applied, in order to preserve the native angular resolution of the data and the intrinsic surface brightness distribution. The green dashed circle marks the extraction region adopted for the compact nuclear core. The blue dashed ellipses indicate the regions associated with Component~A and Component~B of the extended jet, while the cyan dashed ellipse encloses the total large-scale relativistic jet emission. Red crosses denote the photon-weighted centroids of the corresponding components. A $1''$ scale bar is shown for reference. North is up and East is left.}
         \label{fig:X-ray_image}
\end{figure}

The final merged dataset, totaling $360.2\,\mathrm{ksec}$ of effective exposure, yields the deepest X-ray view of PSO~J0309+27 obtained to date and is shown in Fig.~\ref{fig:X-ray_image}. 

Fig.~\ref{fig:X-ray_image} reveals a complex X-ray morphology, characterized by a dominant compact component and additional emission extending on arcsecond scales. Earlier \textit{Chandra} observations of this source, based on a total exposure of $128\,\mathrm{ksec}$, already suggested the presence of extended X-ray emission aligned with the radio jet axis (see \citealt{Ighina2022a}). The increased depth of the present dataset (by a factor $\sim$3) allows a more detailed and quantitative characterization of this structure, enabling a robust investigation of its spatial extent and morphology in the X-ray band. 

Beyond the bright central emission associated with the nucleus, an excess of X-ray photons is detected preferentially along a north-eastern (NE) direction on the plane of the sky. This orientation is consistent with the position angle of the one-sided, parsec-scale radio jet revealed by Very Long Baseline Array (VLBA) observations, which resolve a continuous structure extending up to $\sim 500$ pc in projection \citep{Spingola2020}. The extended X-ray morphology is broadly consistent with the propagation axis of the innermost radio jet.

To quantitatively describe the X-ray morphology highlighted in Fig. \ref{fig:X-ray_image}, we defined a set of extraction regions corresponding to the distinct emission components visible in the image. We chose a 4.5$''\times$1.8$''$ (25$\times$10~kpc) ellipse for the overall jet region  and a $1.5''$ ($\sim8.5$~kpc) circle for component A and B. 
The compact, point-like emission associated with the nucleus is enclosed within a circular aperture with a $2''$ radius, from which a total of 679 photons are detected. 

The large-scale extended emission is sampled by an elliptical region encompassing the full diffuse structure visible on arcsecond scales, yielding a total of 84 detected photons. The local background was estimated from a source-free concentric annulus centered on PSO~J0309+27 and scaled to the source extraction area, resulting in an expected background contribution of $12.7$ photons within the jet region. The observed excess therefore clearly indicates the presence of statistically significant extended X-ray emission.
Within this elliptical aperture, two main regions can be distinguished in the X-ray surface brightness distribution, hereafter referred to as Component~A and Component~B. These regions contain 40 and 29 detected photons, respectively, compared to expected background contributions of only $3.8$ and $3.3$ photons.  

The merged exposure map, accounting for both the effective exposure and effective area, is highly uniform within a $15''\times15''$ region centered on the source. The mean value is $1.12\times10^8\,\mathrm{cm^2\,s}$, with a peak-to-valley variation of $0.12\%$ and a standard deviation of $0.04\%$. These small variations demonstrate that spatial changes in the instrumental sensitivity are negligible across the region considered here and therefore do not significantly affect the morphological analysis.

While the X-ray morphology shown in Fig.~\ref{fig:X-ray_image} suggests the presence of emission extending beyond the compact nuclear component, a quantitative assessment is required to establish whether the detected signal in the jet regions is spatially resolved or consistent with the wings of the instrumental point spread function (PSF). We therefore simulated the expected instrumental PSF using the \texttt{simulate\_psf} tool in \texttt{CIAO}, scaling it to the measured core brightness within a $10\,\mathrm{pixel}$ radius reference aperture. Within the elliptical jet extraction region, the scaled PSF predicts $\sim2.15$ counts, compared to the $84$ observed events, demonstrating that contamination from the nuclear PSF is negligible. To further investigate this, we extracted radial surface brightness profiles from the merged \textit{Chandra} event file in the $0.5$--$7.0\,\mathrm{keV}$ energy band.

For the nuclear component, radial distances were computed with respect to the photon-weighted centroid of the core emission. Surface brightness profiles were then extracted using concentric annuli with a fixed radial width of $0.5''$, computing the background-subtracted photon surface density in each annulus. The background level was estimated from nearby source-free regions of the field and subtracted uniformly from the image prior to profile extraction. 

\begin{figure}[ht!]
   \centering
   \includegraphics[width=\hsize]{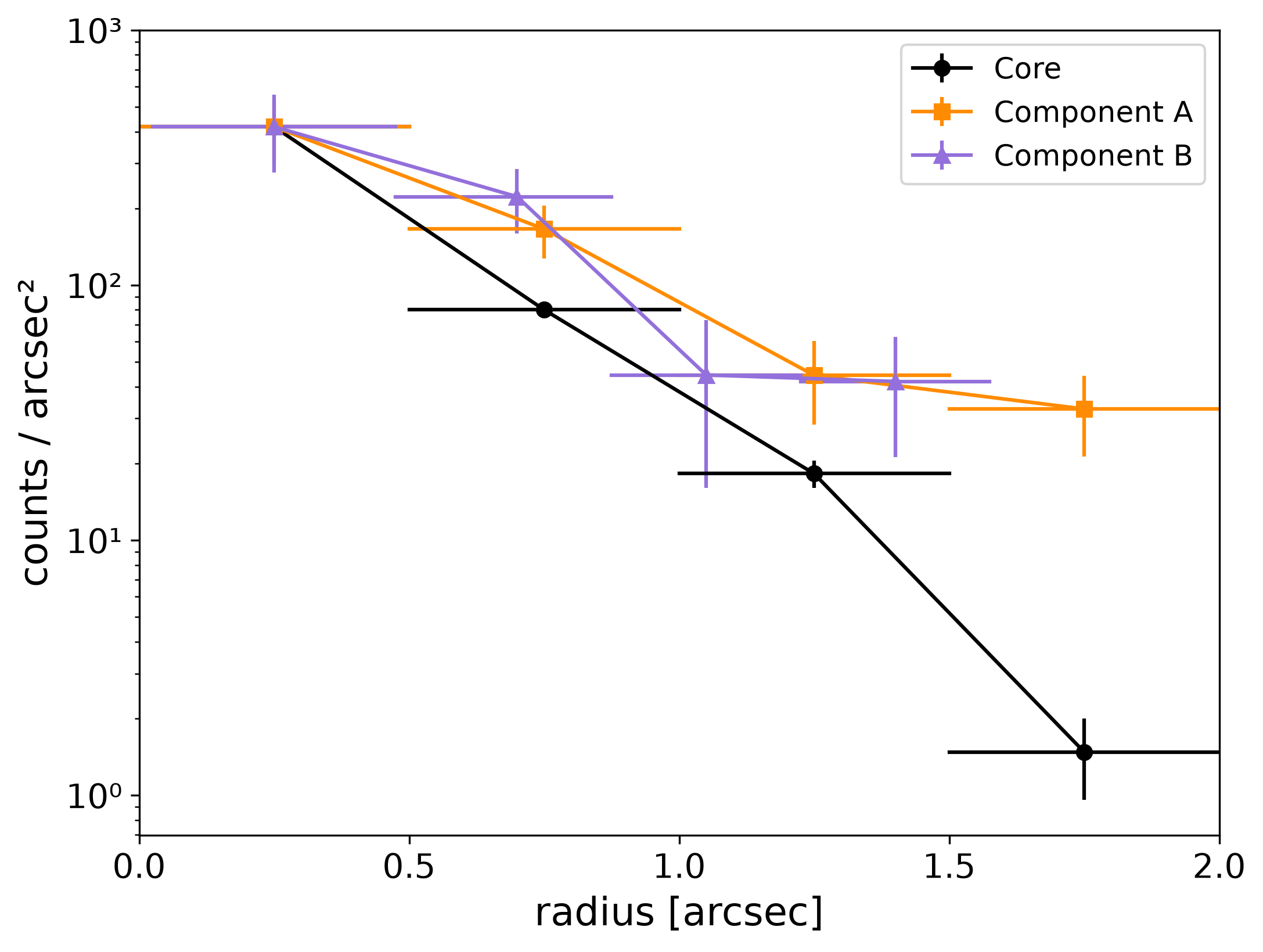}

   \caption{Radial surface brightness profiles of PSO~J0309+27 in the $0.5$--$7.0\,\mathrm{keV}$ energy band. The black curve corresponds to the core, while the orange dashed curve with squares and the light-blue dash-dotted curve with triangles show the profiles extracted for Component~A and Component~B, respectively. The radial binning for each component was chosen to balance the available photon statistics with the spatial sampling of the surface brightness distribution. A constant bin width of $0.5''$ was adopted for the Core and Component~A, while flexible bin sizes of $0.35''$--$0.5''$ were used for Component~B. The jet profiles have been normalized to the innermost bin of the core to facilitate comparison. Horizontal bars indicate the actual radial bin width, while vertical error bars represent $1\sigma$ Poissonian uncertainties.}
   \label{fig:radial_profile}
\end{figure}

An analogous procedure was applied to the extended emission, extracting radial profiles separately for Component A and Component B. In this case, only events within the corresponding circular apertures were considered, and radial distances were measured relative to the photon-weighted centroid of each component. This approach enables a direct comparison between the spatial distribution of the jet emission and that of the compact core, independent of absolute flux normalization. 
The resulting radial surface brightness profiles are shown in Fig.~\ref{fig:radial_profile}. 

The core profile exhibits the steep radial decline expected for a point-like source observed with \textit{Chandra}. In contrast, the profiles associated with Component A and Component B display significantly flatter distributions, with an excess of counts persisting at larger radii relative to the core. This behaviour is not consistent with a purely point-like PSF, and indicates that the emission in the jet regions is spatially extended on arcsecond scales. This result is consistent with earlier indications of a low-surface-brightness extension and a possible change in jet orientation on large scales suggested by \citet{Ighina2022a}, now supported by the deeper dataset.

\section{X-ray spectral analysis} \label{sec:chandra_spectra}
To further characterize the X-ray emission properties of the nuclear and extended components, we performed a spatially resolved spectral analysis of the regions identified in Fig.~\ref{fig:X-ray_image}. Spectra were extracted from the core, Component A, and Component B, as well as from the aperture encompassing the full extended jet structure, adopting the same spatial definitions used in the imaging analysis. 

The spectral extraction procedure follows that adopted for PSO~J0309+27 in previous works (e.g. \citealt{Moretti2021}; \citealt{Ighina2022a}). Source and background spectra were extracted separately for each observation using the \texttt{specextract} task within CIAO, applied to the corresponding reprocessed event files. For the nuclear region, the \texttt{correctpsf} option was enabled in order to account for the encircled energy fraction of the point spread function, while no PSF correction was applied to the extended jet regions. Local background spectra were extracted from nearby source-free regions on the same CCD chip to minimize contamination from diffuse emission and unrelated point sources. The resulting spectra from all observations were then combined using the \texttt{combine\_spectra} script, producing a single spectrum and associated response files for each region.

Spectral modeling was carried out with XSPEC (v12.15.0) in the $0.5$--$7.0\,\mathrm{keV}$ energy range. Given the limited photon statistics in the jet components, we adopted the modified C-statistic (\citealt{Cash1979}), which accounts for a Poisson-distributed background via the W-statistic formalism (\citealt{Wachter1979}). All spectra were modeled with an absorbed power law. The hydrogen column density was fixed to the Galactic value, $N_{\rm H} = 1.13 \times 10^{21}\,\mathrm{cm^{-2}}$, derived from the HI4PI survey (\citealt{HI4PI2016}), while the photon index and normalization were left free to vary. All uncertainties are quoted at the 90\% confidence level.

From the best-fit models we derived the photon index, the observed flux in the $0.5$--$7.0\,\mathrm{keV}$ band, and the intrinsic luminosity in the source rest frame at $z = 6.1$. The resulting spectral parameters are summarized in Table~\ref{table:2}. 
\renewcommand{\arraystretch}{1.5}
\begin{table}[t!]
\centering
\caption{Best-fit spectral parameters for the core and jet components of PSO~J0309+27.}
\label{table:2}
\small 
\begin{tabular}{c c c c}
\hline\hline
\noalign{\smallskip}
Regions & $\Gamma$ & $F_{0.5-7\,\mathrm{keV}}$ & $L_{2-10\,\mathrm{keV}}$ \\
 & & [$10^{-15}$ erg cm$^{-2}$ s$^{-1}$] & [$10^{44}$ erg s$^{-1}$] \\
\noalign{\smallskip}
\hline
\noalign{\smallskip}
Core         & $1.86^{+0.09}_{-0.09}$ & $32.2^{+2.8}_{-2.6}$ & $72.2^{+6.2}_{-5.8}$ \\
Extended Jet & $2.25^{+0.45}_{-0.44}$ & $4.0^{+1.6}_{-1.2}$ & $13.1^{+5.13}_{-4.0}$ \\
Component A  & $2.05^{+0.59}_{-0.56}$ & $2.2^{+1.2}_{-0.9}$ & $6.0^{+3.4}_{-2.3}$ \\
Component B  & $2.54^{+1.13}_{-0.99}$ & $1.4^{+1.5}_{-0.8}$ & $6.0^{+6.4}_{-3.3}$ \\
\noalign{\smallskip}
\hline\hline
\end{tabular}
\tablefoot{The columns report the photon index $\Gamma$, the observed flux in the $0.5$--$7.0\,\mathrm{keV}$ band, and the intrinsic rest-frame $2$--$10\,\mathrm{keV}$ luminosity at $z=6.1$. All fits assume a Galactic-absorbed power-law model with $N_{\rm H}=1.13\times10^{21}\,\mathrm{cm^{-2}}$ (\citealt{HI4PI2016}).}
\end{table}

\begin{figure}[ht!]
    \centering
    \includegraphics[width=\hsize]{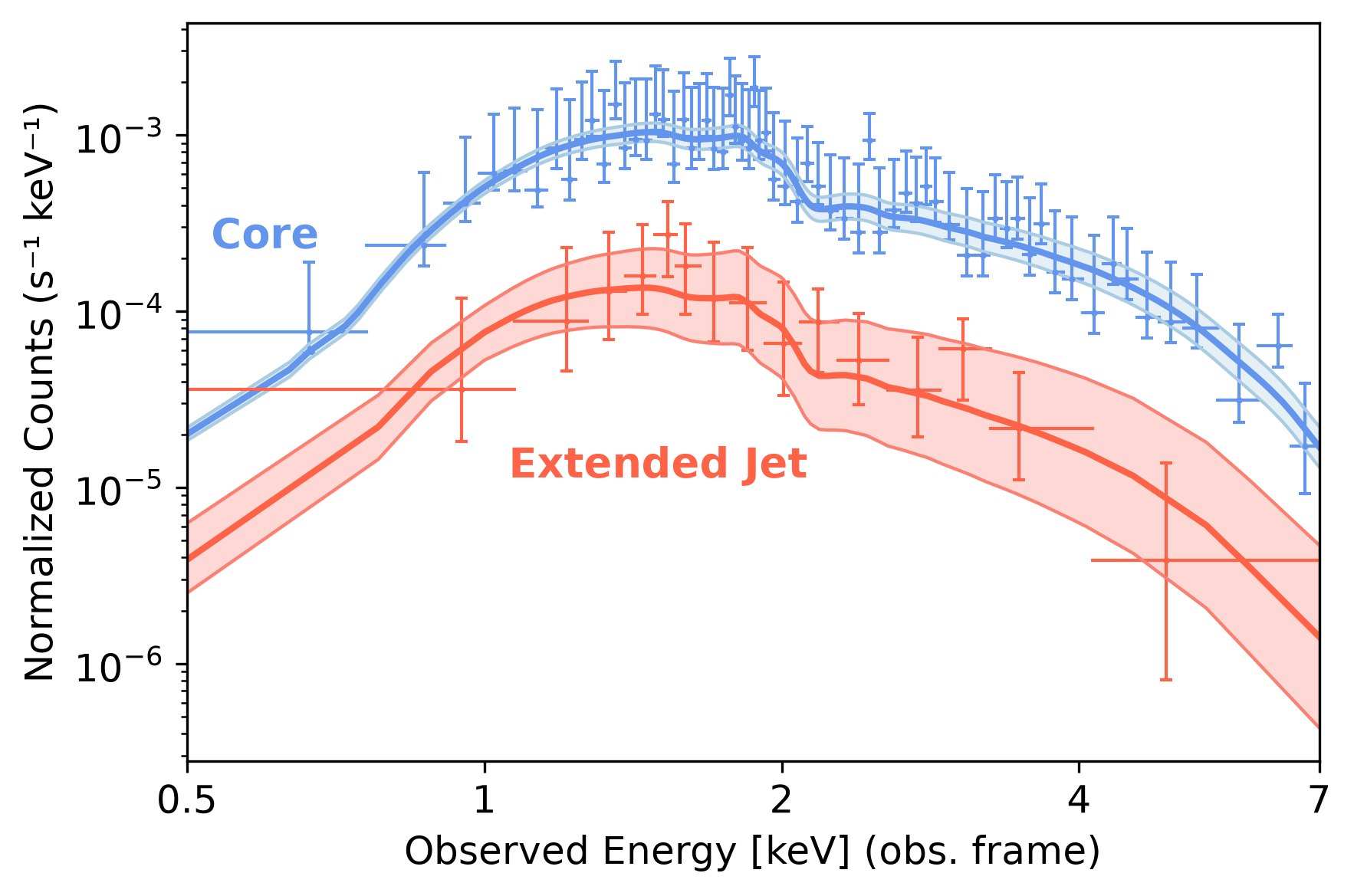}
    \includegraphics[width=\hsize]{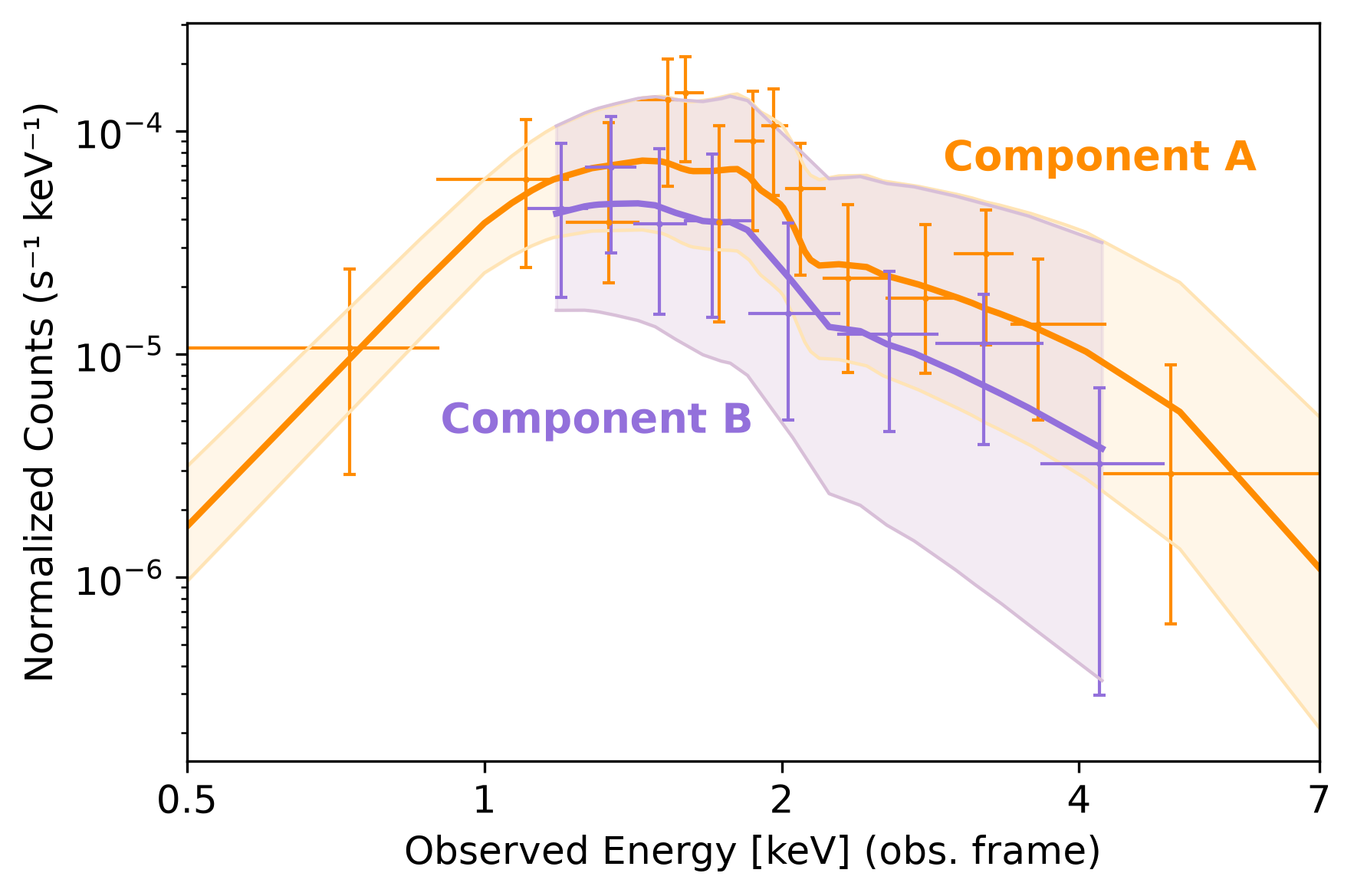}
   
    \caption{Folded \textit{Chandra} X-ray spectra of the core and jet components of PSO~J0309+27 in the $0.5$--$7.0\,\mathrm{keV}$ observed energy range. \textit{Top panel:} nuclear core (blue) and full extended jet (red). \textit{Bottom panel:} Component A (orange) and Component B (purple). Solid lines represent the best-fit Galactic absorbed power-law models, and shaded regions indicate the 90\% confidence intervals. Data are rebinned for display purposes only.}
    \label{fig:spectral_analysis}
\end{figure}

In Fig.~\ref{fig:spectral_analysis} we report the X-ray spectra of the core, Component A, Component B, and the full extended jet structure of PSO~J0309+27 modeled as described above.

The nuclear spectrum is well described by a relatively hard photon index, $\Gamma = 1.86$, consistent with non-thermal emission from a compact and relativistically beamed jet oriented close to the line of sight, as commonly observed in blazar sources (e.g. \citealt{Ghisellini20155}; \citealt{Ighina2019}). The corresponding rest-frame luminosity, $L_{2-10\,\mathrm{keV}} \simeq 7.2 \times 10^{45}\,\mathrm{erg\,s^{-1}}$, confirms the extreme X-ray power of PSO~J0309+27 and places it among the most luminous AGN known at $z>6$ (see, e.g., Fig. 2 in \citealt{Ighina2025b}). Compared to the only other blazar currently known at $z>6$, VLASS J041009.05$-$013919.88 \citep{Banados2024}, PSO~J0309+27 is a about a factor $\sim2$ brighter in the X-rays.

The extended jet component exhibits a softer spectral slope, $\Gamma = 2.25$, suggestive of emission produced by a more evolved population of relativistic electrons on kiloparsec scales. Such softening is commonly observed in large-scale AGN jets and is generally interpreted as the result of radiative cooling, changes in particle acceleration efficiency, or different dominant emission processes with respect to the compact core (e.g. \citealt{Worrall2009}; \citealt{Marshall2018}). The rest-frame luminosity of the extended structure, $L_{2-10\,\mathrm{keV}} \simeq 1.3 \times 10^{45}\,\mathrm{erg\,s^{-1}}$, corresponds to $\sim 18\%$ of the nuclear luminosity, indicating that a non-negligible fraction of the total X-ray output arises from kpc-scale regions.

When considering the two jet components separately, we find $\Gamma = 2.05$ for Component A and $\Gamma = 2.54$ for Component B. Although the uncertainties are large, the data are consistent with a progressive spectral steepening along the jet, which may reflect energy losses of the emitting electrons or structural and dynamical variations within the flow. However, the limited photon statistics prevent a more detailed characterization of the spectral evolution along the jet.

\section{Comparison with the literature}
\label{sec:comparison}

The origin of the high-energy emission observed in kpc-scale jets is still under debate and it is thought to depend on the redshift \citep[e.g.][]{Worrall2020,Tavecchio2021}. On the one hand, at lower redshift the synchrotron emission from a second population of electrons, different from the one responsible for the radio emission, is favored \citep[e.g.][]{Georganopoulos2006,Meyer2014,Breiding2023}. On the other hand, the rapid increase of the CMB energy density, $\propto(1+z)^4$, not only quenches the high-energy synchrotron radiation, but it also boosts the IC/CMB emission \citep[e.g.][]{Ighina2022a}. However, determining the contribution of each mechanism at different redshifts is quite challenging. This is mainly due to the fact that the observed flux depends on many different physical parameters of the jets (e.g. magnetic field, orientation, velocity; \citealt{Worrall2009,Meyer2019,Maithil2025}) and therefore population studies are affected by a large intrinsic scatter. At the same time, the detection of X-ray jets is also challenging from an observational point of view. Given their extended nature on arcsec scales and their relatively faint nature, {\it Chandra} is the only telescope capable of identifying such structures.

\begin{figure}[ht!]
           \centering
           \includegraphics[width=0.9\hsize]{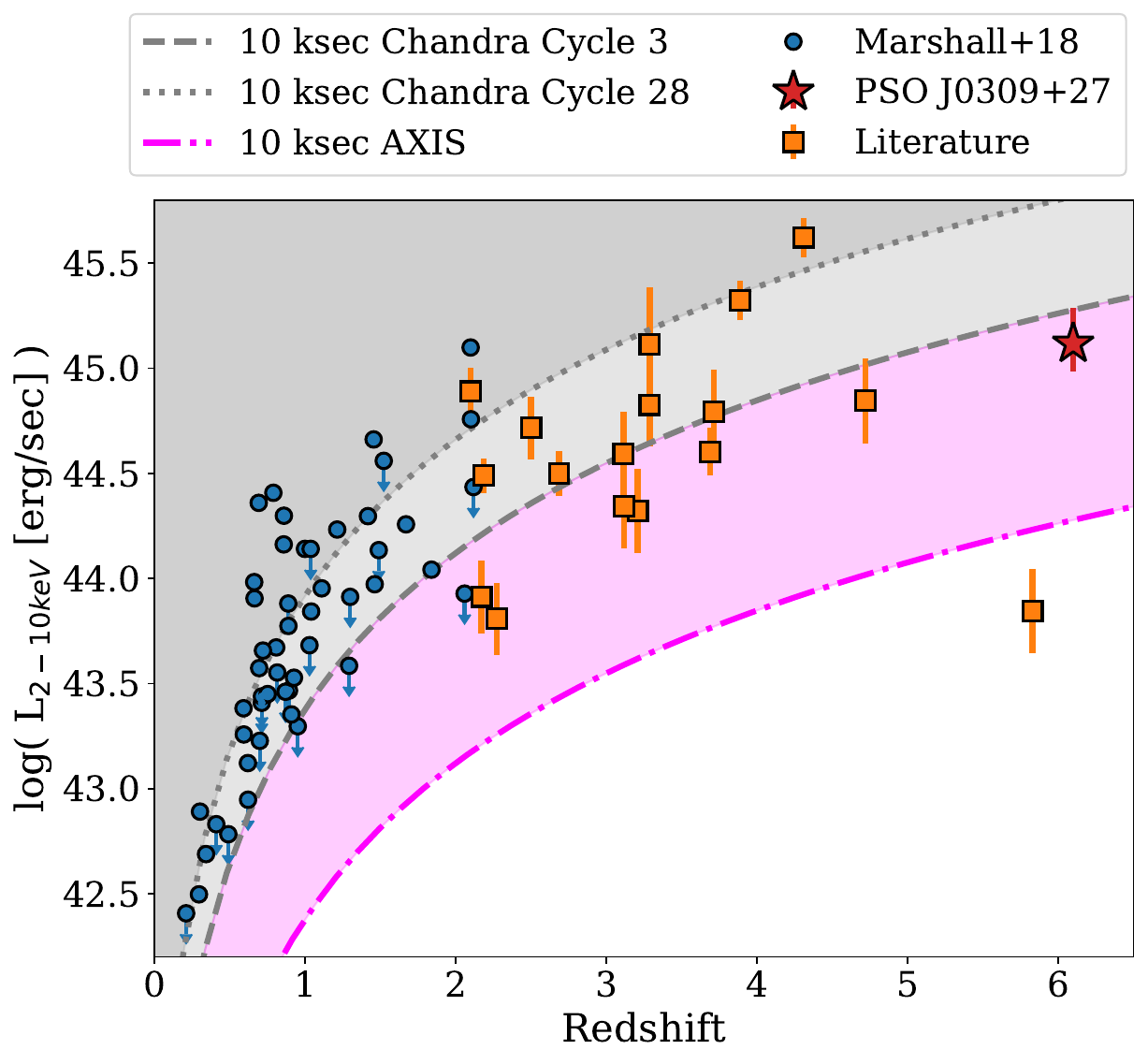}
           \includegraphics[width=0.9\hsize]{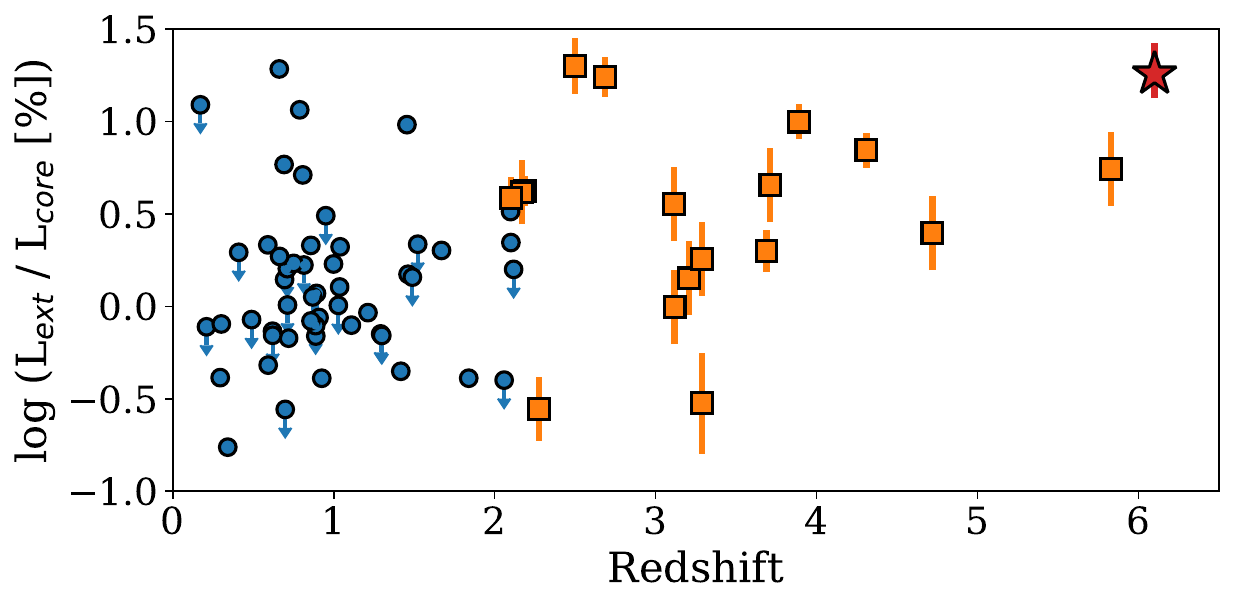}
           \caption{{\it Top panel}: Luminosity in the 2-10~keV rest-frame energy band as a function of redshift of the extended X-ray jets currently known in the literature. Lines and shaded regions show the sensitivity of a 10~ksec exposure with {\it Chandra} at the start of the mission (Cycle 3; dashed gray line) and the current one (Cycle 28; dotted gray line) as well as a 10~ksec exposure with AXIS (dotted-dashed magenta line). {\it Bottom panel}: Ratio of the extended-to-core luminosity ratio for the same jets reported in the above panel.}
         \label{fig:comparison_literature}
\end{figure}

Top panel of Figure \ref{fig:comparison_literature} shows the integrated 2--10~keV rest-frame luminosity of the extended relativistic jets resolved in the literature. These include the survey performed by \cite{Marshall2005,Marshall2011,Marshall2018} with shallow {\it Chandra} observations on $z<2$ extended radio jets ($\sim$6~ksec per target on average). In addition, at $z>2$ we also considered more detailed studies on individual objects with extended X-ray jets (see \citealt{Siemiginowska2003,Cheung2012,Simionescu2016,Schwartz2020,Worrall2020,Snios2021,Connor2021}). The large-scale jet associated with the $z=6.1$ blazar PSO~J0309+27 analyzed in this work corresponds to the most distant kpc-scale jet currently resolved, with only three other similar systems identified at $z>4$ (see \citealt{Siemiginowska2003,Cheung2012,Connor2021}). 
Moreover, the jet associated with PSO~J0309+27 is one of the X-ray brightest jets currently known at all redshifts. 

On average, $z>3$ jets are brighter than the majority of their low-$z$ counterparts from \cite{Marshall2018}. However, this is likely dominated by a selection effect, rather than a redshift evolution due to the IC/CMB process. 
The top panel of Fig. \ref{fig:comparison_literature} shows the sensitivity curve for a relatively short 10~ksec {\it Chandra} exposure to detect a point-like source with $\sim$10 net counts (assuming a photon index of $\Gamma_{\rm X}=2$). Most of the jets currently known lie above the sensitivity curve of {\it Chandra} at the start of operations (Cycle 3, dashed line) since they were discovered from relatively short observations of a large sample \citep[e.g.][]{Marshall2018}.

However, a trend of increasing values with redshift is also potentially present when comparing the extended-to-core X-ray luminosity of jetted AGN (bottom panel in Fig. \ref{fig:comparison_literature}). If no redshift evolution was present in the emission of the extended components, the observed ratio distribution would not change, since most extended jets have been detected close to the detection limit of {\it Chandra}.
For example, in the case of PSO~J0309+27, L$_{\rm ext}/$L$_{\rm core}\approx18\%$, which is larger than most of the X-ray jets currently known at lower redshift. Although a tendency of higher ratio values at higher redshift could in principle be explained by the IC/CMB mechanism, the statistics is still too limited to draw solid conclusions on the presence of an evolution based solely on the extended-to-core ratio. 

In order to have a more representative sample that better describes the true luminosity distribution of these systems at high redshift and perform population evolutionary studies, more sensitive X-ray observations are needed. 
A notable example is the source PSO J352.4034-15.3373 at $z=5.83$, one of the most radio-bright AGN currently known at $z>5$. From a very deep, 265~ksec, {\it Chandra} observation, \cite{Connor2021} uncovered the presence of faint extended X-ray emission in the same direction as the radio jet resolved on VLBI scales \citep{Momjian2018}. As shown in Fig. \ref{fig:comparison_literature}, the detection of this faint component suggests the existence of many more X-ray jets at high redshift that can only be uncovered with deep exposures.  
At the same time, the availability of deep wide-area radio surveys allows for the discovery of many more high-$z$ radio AGN \citep[e.g.][]{Gloudemans2022,Ighina2025,Banados2024} as well as the detection of radio emission from the extended components of their relativistic jets. For example, \cite{Gloudemans2025} recently discovered a powerful radio jet at $z\sim5$ extending about $\sim66$~kpc from the LOw-Frequency ARray (LOFAR) Two-metre Sky Survey (LoTSS; \citealt{Shimwell2019,Shimwell2022}). To perform an efficient and systematic follow-up campaign of the growing number of such systems, a more sensitive X-ray telescope is crucial.
The magenta dotted-dashed line in Fig. \ref{fig:comparison_literature} shows the sensitivity of an X-ray telescope with a collective area $\sim10$ larger than {\it Chandra} capable of resolving X-ray emission on arcsec scales. As detailed in the next section, not only such a new X-ray mission will uncover many more faint X-ray jets in the early Universe, but it will also allow for the detailed characterization, in terms of morphology and spectral properties, of the one already known.

\section{Next-generation X-ray missions}
\label{sec:AXIS}

In the last 25 years the unique angular resolution and sensitivity of the {\it Chandra X-ray observatory} allowed for the discovery of hundreds of extended jets in a wide range of redshift and extension \citep[e.g.][]{Schwartz2000,Harris2006,Worrall2009}.
However, the characterization of AGN at $z\gtrsim6$ remains fundamentally limited by photon statistics and angular resolution constraints of current X-ray facilities, both when studying the emission of their un-resolved core emission (e.g. \citealt{Nanni2017,Vito2019,Ighina2024}) as well as the extended jet emission \citep[e.g.][]{Connor2021,Migliori2023}. In this section, we explore how a potential future X-ray mission with a larger collective area and a similar angular resolution compared to {\it Chandra} will be able to significantly improve studies of high-$z$ jetted AGN. To this end, we use the $z=6.10$ blazar PSO~J0309+27 analyzed in this work as a test case. While our $\sim360\,\mathrm{ksec}$ \textit{Chandra} dataset has successfully resolved the jet morphology, the resulting spectral parameters remain affected by large uncertainties due to the limited number of detected photons. PSO~J0309+27 therefore represents an ideal benchmark for testing the capability of a next-generation X-ray observatory to resolve and characterize large-scale relativistic jets in the early Universe.

To provide concrete estimates and predictions for the possibilities of detecting extended X-ray jets at high redshift, we consider the capabilities proposed for the NASA \textit{Advanced X-ray Imaging Satellite} (AXIS; \citealt{Mushotzky2018}) probe concept. While other mission concepts could provide other vital avenues for studying relativistic jets \citep[e.g.,][]{Marcotulli2024}, the combination of arcsecond angular resolution maintained over a $\sim450$~arcmin$^2$ field of view and an effective area an order of magnitude larger than {\it Chandra} \citep[see][]{Reynolds2024} make the proposed AXIS capabilities the best test case for the potential detection of extended X-ray jets \citep[e.g.][]{Connor2024,Koss2025}, independent of the probe's current status.
The instrumental response files for the specific AXIS mission are publicly available\footnote{\url{https://axis.umd.edu/researchers/simulation-resources}}, including the Redistribution Matrix File (RMF) and the Auxiliary Response File (ARF), which account for energy redistribution and effective area as a function of photon energy. All source and background spectra were folded through these response files to ensure a physically consistent representation of the expected count distribution.

\subsection{Imaging simulations}

To assess the morphological capabilities of AXIS in resolving the kpc-scale structure of PSO~J0309+27, we performed a comprehensive suite of imaging simulations. We adopted a modular approach, independently modeling the astrophysical backgrounds, the nuclear core, and the individual jet components (component A and component B), based on the spectral results obtained from the \textit{Chandra} analysis (see Section~\ref{sec:chandra_spectra}).

The imaging simulation was initialized by constructing the total background, combining the dominant contributions expected for AXIS observations: the Soft X-ray Background (SXRB), the Non–X-ray Background (NXB), and the Cosmic X-ray Background (CXB), following standard spectral prescriptions (e.g. \citealt{McCammon1990}; \citealt{Tawa2008}). Each component was folded through the AXIS response matrices and simulated with an artificially long exposure ($10^8\,\mathrm{s}$) in order to derive stable count rates and suppress stochastic fluctuations.

\begin{figure*}
\centering
\includegraphics[width=\hsize]{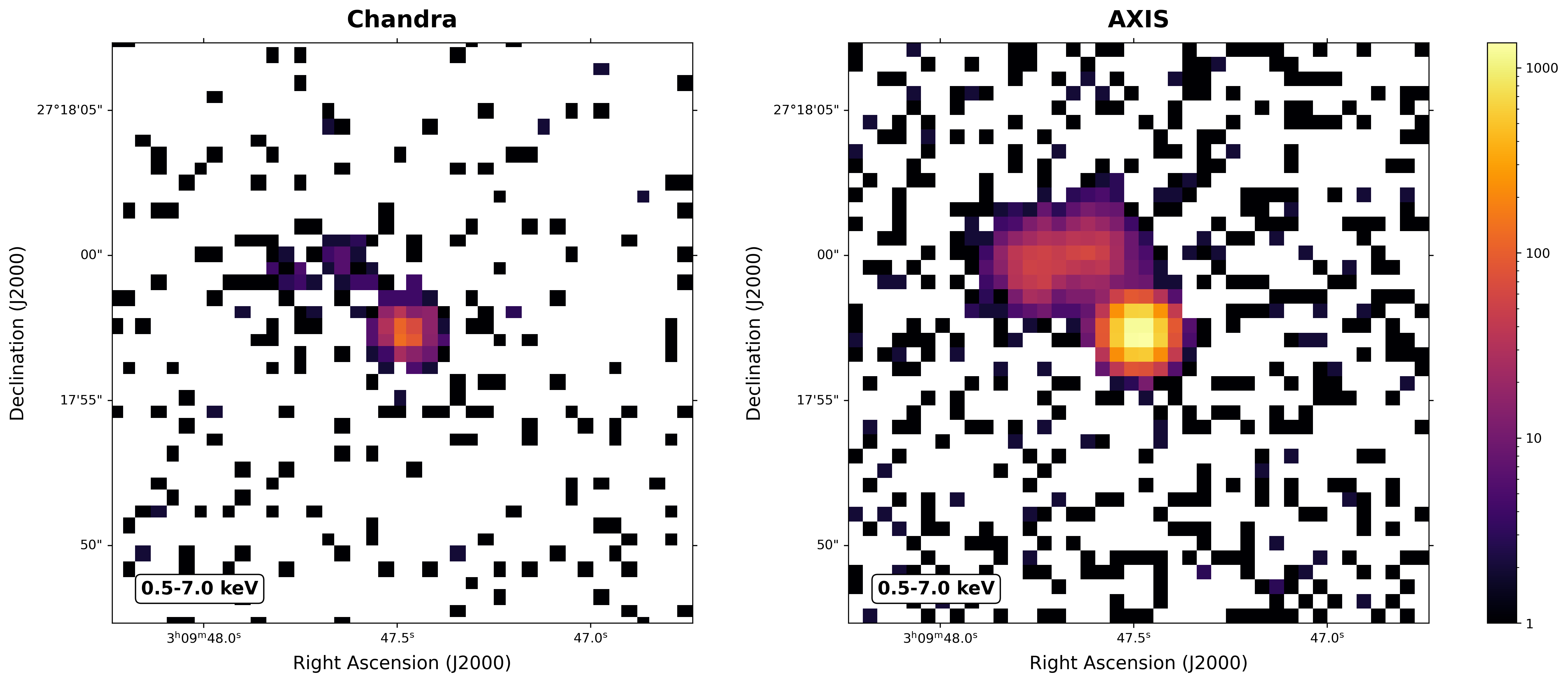}   
    \caption{Comparison between the observed \textit{Chandra} X-ray image (left) and the simulated AXIS mock observation (right) of PSO~J0309+27 in the $0.5$--$7.0\,\mathrm{keV}$ energy band. Both images correspond to an effective exposure of $360.2\,\mathrm{ksec}$ and are displayed using the same logarithmic colour scale.}
    \label{fig:axis_sim}
\end{figure*} 

The resulting background rates were rescaled to the nominal $360\,\mathrm{ksec}$ exposure and combined to obtain the total expected background level. This integrated count rate was then used to populate a two-dimensional grid of $120 \times 120$ pixels, corresponding to a $1 \times 1\,\mathrm{arcmin}^2$ field of view sampled at $0.5\,\mathrm{arcsec}$ per pixel, consistent with the assumed AXIS imaging configuration (\citealt{Reynolds2024}). Poisson realizations were applied on a pixel-by-pixel basis to reproduce the intrinsic counting statistics of an X-ray observation.

The emission from the individual components was modeled as power-law spectra ($F(E) \propto K\,E^{-\Gamma}$), adopting photon indices and normalizations consistent with the \textit{Chandra} spectral analysis (see Table~\ref{table:2}). For each component, the expected count distribution over a $360.2\,\mathrm{ksec}$ exposure was derived by folding the spectral model through the AXIS response matrices within XSPEC. The spatial distribution of the jet emission was modeled using radial surface-brightness profiles derived from the \textit{Chandra} data (see Fig~\ref{fig:radial_profile}), which were used to reconstruct the intrinsic emission maps of the extended components. These maps were subsequently convolved with the AXIS point-spread function (PSF) to account for the instrumental response. The nuclear core was positioned at the center of the field, while the two jet features were placed at the angular separations derived from the {\it Chandra} image. Independent Poisson realizations were applied to the source maps and subsequently combined with the synthetic background image, yielding a complete mock AXIS observation that self-consistently incorporates instrumental response, spatial resolution, and counting statistics.

The resulting simulated image is presented in Figure~\ref{fig:axis_sim}, alongside the corresponding \textit{Chandra} observation. The visual comparison highlights the substantial gain in photon statistics and surface-brightness sensitivity achievable with AXIS.

For an effective exposure of $360\,\mathrm{ksec}$, the simulated AXIS observation predicts an increase in photon statistics across all components. The nuclear core counts rise from $679$ in the \textit{Chandra} data to $1.2\times10^4$ photons in the AXIS simulation, corresponding to an enhancement by a factor of $\sim17$. Similarly, the total detected counts for the extended jet structure increase from 84 to $1.5\times10^3$ photons.
The significantly enhanced photon counts would enable a much cleaner separation of the extended emission from the nuclear component, with reduced contamination from background fluctuations.

\subsection{Simulated AXIS Spectra}
In addition to the imaging analysis, we extended the comparison between \textit{Chandra} and AXIS to the spectral domain, focusing on the extended X-ray jet and its substructures. Simulated AXIS spectra were generated for the jet extraction regions identified in Fig.~\ref{fig:X-ray_image}. Since the nuclear core is already detected with high significance in the existing \textit{Chandra} observations, the comparison presented here is restricted to the extended components, where the improved collecting area of AXIS is expected to provide the largest scientific gain.

\begin{figure*}
\centering
\includegraphics[width=\hsize]{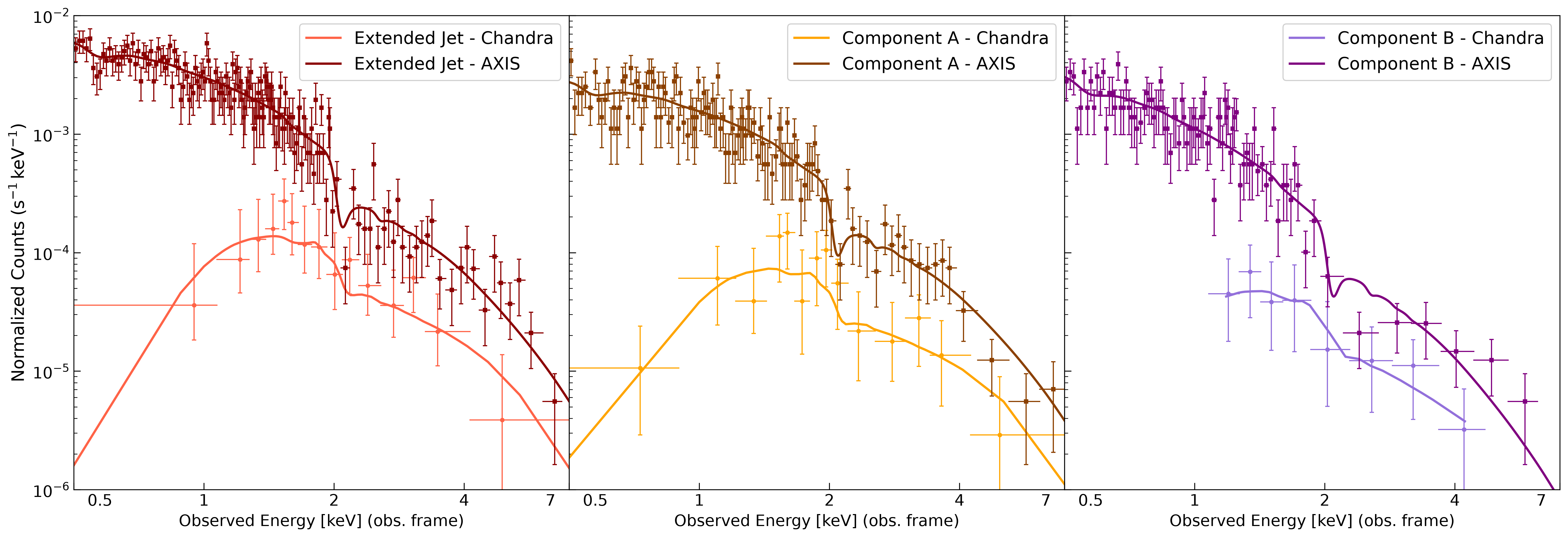}
    \caption{Comparison between the observed \textit{Chandra} spectra and the simulated AXIS spectra of PSO~J0309+27 in the $0.5$--$7.0\,\mathrm{keV}$ band. The panels show the extended jet (left), Component A (center), and Component B (right). Lighter symbols correspond to \textit{Chandra}, darker symbols to AXIS. Data points are rebinned for visualization purposes, while solid lines represent the best-fitting models folded through the respective instrumental responses.}
    \label{fig:axis_spectra}
\end{figure*}

The simulations were performed within the XSPEC environment using the \texttt{fakeit} command. For each region, the best-fitting spectral models derived from the \textit{Chandra} data (see Table~\ref{table:2}), were used as input and folded through the official AXIS Phase~A response matrices (RMF and ARF). A synthetic spectrum was then generated assuming an exposure time of $360\,\mathrm{ksec}$, matching the existing \textit{Chandra} dataset. This procedure ensures a direct, one-to-one comparison between the observed \textit{Chandra} spectra and the expected AXIS performance under identical observing conditions, isolating the impact of instrumental sensitivity and response.
The resulting simulated spectra are shown in Fig.~\ref{fig:axis_spectra}.

Although the intrinsic spectral models are identical by construction, the comparison in Figure~\ref{fig:axis_spectra} illustrates the impact of observing the same emission through the AXIS instrumental response. The larger effective area, particularly in the soft band, produces a substantial increase in detected photon counts across all regions. This enhancement translates into a higher normalization in count space, a denser sampling of the $0.5$--$7.0\,\mathrm{keV}$ energy range, and significantly reduced statistical uncertainties on individual spectral bins. As a result, the response-convolved spectral shape appears more clearly defined, with tighter constraints on the power-law slope and normalization.

For the jet components, the transition from a few tens of photons in the \textit{Chandra} data to several hundred counts in the AXIS simulation moves the analysis from a photon-starved regime to one in which independent spectral fitting becomes statistically robust. The improvement is therefore not merely quantitative but qualitative, enabling significantly tighter parameter constraints and reliable component separation. Monte Carlo simulations of the AXIS spectra (using the \texttt{fakeit} task in \texttt{XSPEC}) indicate that the uncertainty on the photon index of the extended jet decreases from the current $\Delta\Gamma\sim0.45$ measured with \textit{Chandra} (Table~\ref{table:2}) to an expected $\Delta\Gamma\sim0.15$ for a $360\,\mathrm{ksec}$ AXIS observation. Similar improvements are found for the individual jet substructures, with the uncertainty decreasing from $\sim0.6$ to $\sim0.24$ for Component~A and from $\sim1.1$ to $\sim0.40$ for Component~B. Reducing uncertainties on the photon index of extended jets will be crucial in order to understand the origin of their high-energy emission. One of the main predictions of the IC/CMB mechanism is that the slope of the X-ray spectrum matches the one observed at lower frequencies, in the radio band, since they are both produced by the same population of electrons \citep[e.g.][]{Worrall2009}. However, this  could not be tested for the vast majority of extended jets currently known, due to the low photon statistics and, as a consequence, the large uncertainties on the photon index. A new generation X-ray survey will provide constraints tight enough to be compared to the radio spectral index and test the IC/CMB scenario for a statistical number of systems. 

\section{Summary and conclusions}
\label{sec:conclusions}

In this work, we have presented the deepest X-ray view to date of the $z=6.10$ blazar PSO~J0309+27, based on a total of 22 \textit{Chandra} exposures combining archival and proprietary data for a cumulative exposure of $360\,\mathrm{ksec}$. This dataset represents one of the most sensitive X-ray investigations of a jetted AGN in the early Universe, enabling the detailed study of its morphology and spectral properties.

The imaging analysis reveals a complex X-ray morphology where the dominant nuclear emission is accompanied by a statistically significant extended structure. By leveraging the depth of our dataset, we successfully resolve this emission into two distinct features, Component A and Component B, aligned with the known radio jet axis. We detect 679 photons from the compact core and a total of 84 photons from the extended jet, confirming and substantially strengthening previous indications of large-scale emission (\citealt{Ighina2022a}). This interpretation is quantitatively supported by our radial surface brightness analysis: the jet components exhibit a significantly broader distribution than the instrumental PSF, providing robust evidence for a spatially resolved kpc-scale jet at $z>6$.

Based on the spectral analysis of the different components, the nuclear emission is well described by a hard power-law spectrum ($\Gamma = 1.86^{+0.09}_{-0.09}$), consistent with a relativistically beamed blazar core \citep[e.g.][]{Ghisellini2015b}, while the extended jet exhibits a significantly softer slope ($\Gamma = 2.25^{+0.45}_{-0.44}$). Although limited by photon statistics, these measurements are not sufficient to discriminate between the different emission mechanisms proposed for kpc-scale jets. In particular, while synchrotron emission is generally favored at low redshift, the $(1+z)^4$ increase of the CMB energy density at $z>6$ is expected to enhance the IC/CMB contribution, making the interpretation of the observed X-ray emission highly degenerate (\citealt{Tavecchio2021}). In this context, the detection and characterization of resolved X-ray jets at such high redshift remain both observationally challenging and crucial to constrain the physical processes governing relativistic outflows in the early Universe.

Placing these results in a broader context, PSO~J0309+27 represents the most distant kpc-scale X-ray jet currently resolved, pushing the observational frontier of jetted AGN studies into the epoch of reionization. The extended X-ray luminosity and the extended-to-core luminosity ratio ($L_{\rm ext}/L_{\rm core} \sim 18\%$) are higher than most extended systems observed in lower-redshift systems. While this could be interpreted as evidence for an enhanced IC/CMB contribution at high redshift, current samples remain strongly affected by observational biases, preventing robust conclusions on the redshift evolution of jet emission processes. 

In this context, the combination of limited photon statistics and strong selection biases defines the current observational ceiling of the high redshift X-ray astronomy. Overcoming these limitations requires a new generation of high-resolution observatories. Within this framework, we have performed simulations based on the AXIS mission concept (\citealt{Reynolds2024}), which combines an order-of-magnitude increase in sensitivity with \textit{Chandra}-like angular resolution. Our results show that a mission with similar specifications would enable both the detection of a fainter population of high-redshift jets and a precise spectral characterization of AGN feedback across cosmic time (see also \citealt{Koss2025}). 

Ultimately, the case of PSO~J0309+27 serves as a benchmark, highlighting that the future of X-ray astronomy lies in the transition from the discovery of individual extreme objects to performing statistically robust population studies at the edge of the observable Universe.

\begin{acknowledgements}
The authors would like to thank J. Maithil, D. Schwartz and A. Siemiginowska for helpful discussions.\\
F.A.M., L.I., A. C. and A. M. acknowledge financial support from INAF under the projects “Quasar jets in the early Universe” (Ricerca Fondamentale 2022) and “Testing the obscuration in the early Universe” (Ricerca Fondamentale 2023).\\
Support for L.I.'s  work was provided by the National Aeronautics and Space Administration through Chandra Award Number GO3-24069X issued by the {\it Chandra X-ray Observatory Center}, which is operated by the Smithsonian Astrophysical Observatory for and on behalf of the National Aeronautics Space Administration under contract NAS8-03060.
T.C. acknowledges support from NASA Contract NAS8-03060 to the \textit{Chandra} X-ray Center.\\
C.~S. acknowledges financial support by the Italian Ministry of University and Research (grant FIS2023$-$01611, CUP C53C25000300001) and by the National Institute for Astrophysics (INAF) $-$ Ricerca Fondamentale 2024 (Ob.~Fu. 1.05.24.07.04).\\
The scientific results reported in this article are based on observations made by the \textit{Chandra X-ray Observatory} contained in the \textit{Chandra} Data Collection (CDC) 587~\href{https://doi.org/10.25574/cdc.587}{doi:10.25574/cdc.587}.

\end{acknowledgements}
\bibliographystyle{bibtex/aa} 
\bibliography{References}

\begin{thebibliography}{66}
\expandafter\ifx\csname natexlab\endcsname\relax\def\natexlab#1{#1}\fi

\bibitem[{{Ba{\~n}ados} {et~al.}(2025){Ba{\~n}ados}, {Momjian}, {Connor}, {Belladitta}, {Decarli}, {Mazzucchelli}, {Venemans}, {Walter}, {Wang}, {Xie}, {Barth}, {Eilers}, {Fan}, {Khusanova}, {Schindler}, {Stern}, {Yang}, {Andika}, {Carilli}, {Farina}, {Fabian}, {Hennawi}, {Pensabene}, \& {Rojas-Ruiz}}]{Banados2024}
{Ba{\~n}ados}, E., {Momjian}, E., {Connor}, T., {et~al.} 2025, Nature Astronomy

\bibitem[{{Belladitta} {et~al.}(2020){Belladitta}, {Moretti}, {Caccianiga}, {Spingola}, {Severgnini}, {Della Ceca}, {Ghisellini}, {Dallacasa}, {Sbarrato}, {Cicone}, {Cassar{\`a}}, \& {Pedani}}]{Belladitta2020}
{Belladitta}, S., {Moretti}, A., {Caccianiga}, A., {et~al.} 2020, \aap, 635, L7

\bibitem[{{Blandford} {et~al.}(2019){Blandford}, {Meier}, \& {Readhead}}]{Blandford2019}
{Blandford}, R., {Meier}, D., \& {Readhead}, A. 2019, \araa, 57, 467

\bibitem[{{Breiding} {et~al.}(2017){Breiding}, {Meyer}, {Georganopoulos}, {Keenan}, {DeNigris}, \& {Hewitt}}]{Breiding2017}
{Breiding}, P., {Meyer}, E.~T., {Georganopoulos}, M., {et~al.} 2017, \apj, 849, 95

\bibitem[{{Breiding} {et~al.}(2023){Breiding}, {Meyer}, {Georganopoulos}, {Reddy}, {Kollmann}, \& {Roychowdhury}}]{Breiding2023}
{Breiding}, P., {Meyer}, E.~T., {Georganopoulos}, M., {et~al.} 2023, \mnras, 518, 3222

\bibitem[{{Cash}(1979)}]{Cash1979}
{Cash}, W. 1979, \apj, 228, 939

\bibitem[{{Celotti} {et~al.}(2001){Celotti}, {Ghisellini}, \& {Chiaberge}}]{Celotti2001}
{Celotti}, A., {Ghisellini}, G., \& {Chiaberge}, M. 2001, \mnras, 321, L1

\bibitem[{Chambers {et~al.}(2016)Chambers, Magnier, Metcalfe, Flewelling, Huber, Waters, Denneau, Draper, Farrow, Finkbeiner, Holmberg, Koppenhoefer, Price, Rest, Saglia, Schlafly, Smartt, Sweeney, Wainscoat, Burgett, Chastel, Grav, Heasley, Hodapp, Jedicke, Kaiser, Kudritzki, Luppino, Lupton, Monet, Morgan, Onaka, Shiao, Stubbs, Tonry, White, Ba{\~{n}}ados, Bell, Bender, Bernard, Boegner, Boffi, Botticella, Calamida, Casertano, Chen, Chen, Cole, Deacon, Frenk, Fitzsimmons, Gezari, Gibbs, Goessl, Goggia, Gourgue, Goldman, Grant, Grebel, Hambly, Hasinger, Heavens, Heckman, Henderson, Henning, Holman, Hopp, Ip, Isani, Jackson, Keyes, Koekemoer, Kotak, Le, Liska, Long, Lucey, Liu, Martin, Masci, McLean, Mindel, Misra, Morganson, Murphy, Obaika, Narayan, Nieto-Santisteban, Norberg, Peacock, Pier, Postman, Primak, Rae, Rai, Riess, Riffeser, Rix, R{\"{o}}ser, Russel, Rutz, Schilbach, Schultz, Scolnic, Strolger, Szalay, Seitz, Small, Smith, Soderblom, Taylor, Thomson, Taylor, Thakar, Thiel, Thilker, Unger, Urata,
  Valenti, Wagner, Walder, Walter, Watters, Werner, Wood-Vasey, \& Wyse}]{Chambers2016}
Chambers, K.~C., Magnier, E.~A., Metcalfe, N., {et~al.} 2016, eprint arXiv:1612.05560 [\eprint[arXiv]{1612.05560}]

\bibitem[{{Cheung} {et~al.}(2012){Cheung}, {Stawarz}, {Siemiginowska}, {Gobeille}, {Wardle}, {Harris}, \& {Schwartz}}]{Cheung2012}
{Cheung}, C.~C., {Stawarz}, {\L}., {Siemiginowska}, A., {et~al.} 2012, \apjl, 756, L20

\bibitem[{{Connor} {et~al.}(2024){Connor}, {Ba{\~n}ados}, {Cappelluti}, \& {Foord}}]{Connor2024}
{Connor}, T., {Ba{\~n}ados}, E., {Cappelluti}, N., \& {Foord}, A. 2024, Universe, 10, 227

\bibitem[{{Connor} {et~al.}(2021){Connor}, {Ba{\~n}ados}, {Stern}, {Carilli}, {Fabian}, {Momjian}, {Rojas-Ruiz}, {Decarli}, {Farina}, {Mazzucchelli}, \& {Earnshaw}}]{Connor2021}
{Connor}, T., {Ba{\~n}ados}, E., {Stern}, D., {et~al.} 2021, \apj, 911, 120

\bibitem[{{Diana} {et~al.}(2022){Diana}, {Caccianiga}, {Ighina}, {Belladitta}, {Moretti}, \& {Della Ceca}}]{Diana2022}
{Diana}, A., {Caccianiga}, A., {Ighina}, L., {et~al.} 2022, \mnras, 511, 5436

\bibitem[{{Fabian}(2012)}]{Fabian2012}
{Fabian}, A.~C. 2012, \araa, 50, 455

\bibitem[{{Fruscione} {et~al.}(2006){Fruscione}, {McDowell}, {Allen}, {Brickhouse}, {Burke}, {Davis}, {Durham}, {Elvis}, {Galle}, {Harris}, {Huenemoerder}, {Houck}, {Ishibashi}, {Karovska}, {Nicastro}, {Noble}, {Nowak}, {Primini}, {Siemiginowska}, {Smith}, \& {Wise}}]{Fruscinone2006}
{Fruscione}, A., {McDowell}, J.~C., {Allen}, G.~E., {et~al.} 2006, SPIE, 6270, 62701V

\bibitem[{{Garmire} {et~al.}(2003){Garmire}, {Bautz}, {Ford}, {Nousek}, \& {Ricker}}]{Garmire2003}
{Garmire}, G.~P., {Bautz}, M.~W., {Ford}, P.~G., {Nousek}, J.~A., \& {Ricker}, George~R., J. 2003, 4851, 28

\bibitem[{{Georganopoulos} {et~al.}(2006){Georganopoulos}, {Perlman}, {Kazanas}, \& {McEnery}}]{Georganopoulos2006}
{Georganopoulos}, M., {Perlman}, E.~S., {Kazanas}, D., \& {McEnery}, J. 2006, \apjl, 653, L5

\bibitem[{{Ghisellini}(2015)}]{Ghisellini20155}
{Ghisellini}, G. 2015, Journal of High Energy Astrophysics, 7, 163

\bibitem[{{Ghisellini} {et~al.}(2010){Ghisellini}, {Della Ceca}, {Volonteri}, {Ghirlanda}, {Tavecchio}, {Foschini}, {Tagliaferri}, {Haardt}, {Pareschi}, \& {Grindlay}}]{Ghisellini2010}
{Ghisellini}, G., {Della Ceca}, R., {Volonteri}, M., {et~al.} 2010, \mnras, 405, 387

\bibitem[{{Ghisellini} {et~al.}(2015){Ghisellini}, {Tagliaferri}, {Sbarrato}, \& {Gehrels}}]{Ghisellini2015b}
{Ghisellini}, G., {Tagliaferri}, G., {Sbarrato}, T., \& {Gehrels}, N. 2015, \mnras, 450, L34

\bibitem[{{Gloudemans} {et~al.}(2022){Gloudemans}, {Duncan}, {Saxena}, {Harikane}, {Hill}, {Zeimann}, {R{\"o}ttgering}, {Yang}, {Best}, {Ba{\~n}ados}, {Drabent}, {Hardcastle}, {Hennawi}, {Lansbury}, {Magliocchetti}, {Miley}, {Nanni}, {Shimwell}, {Smith}, {Venemans}, \& {Wagenveld}}]{Gloudemans2022}
{Gloudemans}, A.~J., {Duncan}, K.~J., {Saxena}, A., {et~al.} 2022, \aap, 668, A27

\bibitem[{{Gloudemans} {et~al.}(2025){Gloudemans}, {Sweijen}, {Morabito}, {Farina}, {Duncan}, {Harikane}, {R{\"o}ttgering}, {Saxena}, \& {Schindler}}]{Gloudemans2025}
{Gloudemans}, A.~J., {Sweijen}, F., {Morabito}, L.~K., {et~al.} 2025, \apjl, 980, L8

\bibitem[{{Harris} \& {Krawczynski}(2006)}]{Harris2006}
{Harris}, D.~E. \& {Krawczynski}, H. 2006, \araa, 44, 463

\bibitem[{{HI4PI Collaboration} {et~al.}(2016){HI4PI Collaboration}, {Ben Bekhti}, {Fl{\"o}er}, {Keller}, {Kerp}, {Lenz}, {Winkel}, {Bailin}, {Calabretta}, {Dedes}, {Ford}, {Gibson}, {Haud}, {Janowiecki}, {Kalberla}, {Lockman}, {McClure-Griffiths}, {Murphy}, {Nakanishi}, {Pisano}, \& {Staveley-Smith}}]{HI4PI2016}
{HI4PI Collaboration}, {Ben Bekhti}, N., {Fl{\"o}er}, L., {et~al.} 2016, Astronomy \& Astrophysics, 594, A116

\bibitem[{{Ighina} {et~al.}(2025{\natexlab{a}}){Ighina}, {Caccianiga}, {Connor}, {Moretti}, {Pacucci}, {Reynolds}, {Afonso}, {Arsioli}, {Belladitta}, {Broderick}, {Dallacasa}, {Della Ceca}, {Haardt}, {Lambrides}, {Leung}, {Lupi}, {Matute}, {Rigamonti}, {Severgnini}, {Seymour}, {Tavecchio}, \& {Vignali}}]{Ighina2025b}
{Ighina}, L., {Caccianiga}, A., {Connor}, T., {et~al.} 2025{\natexlab{a}}, \apjl, 990, L56

\bibitem[{{Ighina} {et~al.}(2019){Ighina}, {Caccianiga}, {Moretti}, {Belladitta}, {Della Ceca}, {Ballo}, \& {Dallacasa}}]{Ighina2019}
{Ighina}, L., {Caccianiga}, A., {Moretti}, A., {et~al.} 2019, \mnras, 489, 2732

\bibitem[{{Ighina} {et~al.}(2021){Ighina}, {Caccianiga}, {Moretti}, {Belladitta}, {Della Ceca}, \& {Diana}}]{Ighina2021b}
{Ighina}, L., {Caccianiga}, A., {Moretti}, A., {et~al.} 2021, \mnras, 505, 4120

\bibitem[{{Ighina} {et~al.}(2024){Ighina}, {Caccianiga}, {Moretti}, {Broderick}, {Leung}, {Paterson}, {Rigamonti}, {Seymour}, {Belladitta}, {Drouart}, {Galvin}, \& {Hurley-Walker}}]{Ighina2024}
{Ighina}, L., {Caccianiga}, A., {Moretti}, A., {et~al.} 2024, \aap, 687, A242

\bibitem[{{Ighina} {et~al.}(2025{\natexlab{b}}){Ighina}, {Caccianiga}, {Moretti}, {Broderick}, {Leung}, {Rigamonti}, {Seymour}, {Afonso}, {Connor}, {Vignali}, {Wang}, {An}, {Arsioli}, {Bisogni}, {Dallacasa}, {Della Ceca}, {Liu}, {L{\'o}pez-S{\'a}nchez}, {Matute}, {Reynolds}, {Rossi}, {Spingola}, {Severgnini}, \& {Tavecchio}}]{Ighina2025}
{Ighina}, L., {Caccianiga}, A., {Moretti}, A., {et~al.} 2025{\natexlab{b}}, \aap, 698, A158

\bibitem[{{Ighina} {et~al.}(2022){Ighina}, {Moretti}, {Tavecchio}, {Caccianiga}, {Belladitta}, {Dallacasa}, {Della Ceca}, {Sbarrato}, \& {Spingola}}]{Ighina2022a}
{Ighina}, L., {Moretti}, A., {Tavecchio}, F., {et~al.} 2022, \aap, 659, A93

\bibitem[{{Koss} {et~al.}(2025){Koss}, {Aftab}, {Allen}, {Amato}, {An}, {Andreoni}, {Anguita}, {Arcodia}, {Ayres}, {Bachetti}, {Baglio}, \& {Bahramian}}]{Koss2025}
{Koss}, M., {Aftab}, N., {Allen}, S.~W., {et~al.} 2025, arXiv e-prints, arXiv:2511.00253

\bibitem[{{Liniewicz} {et~al.}(2026){Liniewicz}, {Stawarz}, {Cheung}, {Migliori}, \& {Siemiginowska}}]{Liniewicz2026}
{Liniewicz}, P., {Stawarz}, {\L}., {Cheung}, C.~C., {Migliori}, G., \& {Siemiginowska}, A. 2026, \apj, 1007, 111

\bibitem[{{Lister} {et~al.}(2019){Lister}, {Homan}, {Hovatta}, {Kellermann}, {Kiehlmann}, {Kovalev}, {Max-Moerbeck}, {Pushkarev}, {Readhead}, {Ros}, \& {Savolainen}}]{Lister2019}
{Lister}, M.~L., {Homan}, D.~C., {Hovatta}, T., {et~al.} 2019, \apj, 874, 43

\bibitem[{{Maithil} {et~al.}(2025){Maithil}, {Schwartz}, {Siemiginowska}, {Kharb}, {Worrall}, {Wardle}, {Migliori}, {Cheung}, {Snios}, {Gobeille}, {Marshall}, \& {Birkinshaw}}]{Maithil2025}
{Maithil}, J., {Schwartz}, D.~A., {Siemiginowska}, A., {et~al.} 2025, \apj, 986, 75

\bibitem[{{Marcotulli} {et~al.}(2024){Marcotulli}, {Ajello}, {B{\"o}ttcher}, {Coppi}, {Costamante}, {Di Gesu}, {Errando}, {Garc{\'\i}a}, {Gokus}, {Liodakis}, {Madejski}, {Madsen}, {Moretti}, {Middei}, {McBride}, {Petropoulou}, {Rani}, {Sbarrato}, {Stern}, {Vasilopoulos}, {Zacharias}, \& {Zhang}}]{Marcotulli2024}
{Marcotulli}, L., {Ajello}, M., {B{\"o}ttcher}, M., {et~al.} 2024, Frontiers in Astronomy and Space Sciences, 11, 1290057

\bibitem[{Marshall {et~al.}(2011)Marshall, {Gelbord}, {Schwartz}, {Murphy}, {Lovell}, {Worrall}, {Birkinshaw}, {Perlman}, {Godfrey}, \& {Jauncey}}]{Marshall2011}
Marshall, H.~L., {Gelbord}, J.~M., {Schwartz}, D.~A., {et~al.} 2011, \apjs, 193, 15

\bibitem[{Marshall {et~al.}(2018)Marshall, Gelbord, Worrall, Birkinshaw, Schwartz, Jauncey, Griffiths, Murphy, Lovell, Perlman, \& Godfrey}]{Marshall2018}
Marshall, H.~L., Gelbord, J.~M., Worrall, D.~M., {et~al.} 2018, \apj, 856, 66

\bibitem[{Marshall {et~al.}(2005)Marshall, {Schwartz}, {Lovell}, {Murphy}, {Worrall}, {Birkinshaw}, {Gelbord}, {Perlman}, \& {Jauncey}}]{Marshall2005}
Marshall, H.~L., {Schwartz}, D.~A., {Lovell}, J.~E.~J., {et~al.} 2005, \apjs, 156, 13

\bibitem[{{McCammon} \& {Sanders}(1990)}]{McCammon1990}
{McCammon}, D. \& {Sanders}, W.~T. 1990, \araa, 28, 657

\bibitem[{{Meyer} \& {Georganopoulos}(2014)}]{Meyer2014}
{Meyer}, E.~T. \& {Georganopoulos}, M. 2014, \apjl, 780, L27

\bibitem[{{Meyer} {et~al.}(2019){Meyer}, {Iyer}, {Reddy}, {Georganopoulos}, {Breiding}, \& {Keenan}}]{Meyer2019}
{Meyer}, E.~T., {Iyer}, A.~R., {Reddy}, K., {et~al.} 2019, \apjl, 883, L2

\bibitem[{{Meyer} {et~al.}(2023){Meyer}, {Shaik}, {Tang}, {Reid}, {Reddy}, {Breiding}, {Georganopoulos}, {Chiaberge}, {Perlman}, {Clautice}, {Sparks}, {DeNigris}, \& {Trevor}}]{Meyer2023}
{Meyer}, E.~T., {Shaik}, A., {Tang}, Y., {et~al.} 2023, Nature Astronomy, 7, 967

\bibitem[{{Migliori} {et~al.}(2023){Migliori}, {Siemiginowska}, {Sobolewska}, {Cheung}, {Stawarz}, {Schwartz}, {Snios}, {Saxena}, \& {Kashyap}}]{Migliori2023}
{Migliori}, G., {Siemiginowska}, A., {Sobolewska}, M., {et~al.} 2023, \mnras, 524, 1087

\bibitem[{{Momjian} {et~al.}(2018){Momjian}, {Carilli}, {Ba{\~n}ados}, {Walter}, \& {Venemans}}]{Momjian2018}
{Momjian}, E., {Carilli}, C.~L., {Ba{\~n}ados}, E., {Walter}, F., \& {Venemans}, B.~P. 2018, \apj, 861, 86

\bibitem[{{Moretti} {et~al.}(2021){Moretti}, {Ghisellini}, {Caccianiga}, {Belladitta}, {Della Ceca}, {Ighina}, {Sbarrato}, {Severgnini}, \& {Spingola}}]{Moretti2021}
{Moretti}, A., {Ghisellini}, G., {Caccianiga}, A., {et~al.} 2021, \apj, 920, 15

\bibitem[{{Mushotzky}(2018)}]{Mushotzky2018}
{Mushotzky}, R. 2018, in Society of Photo-Optical Instrumentation Engineers (SPIE) Conference Series, Vol. 10699, Space Telescopes and Instrumentation 2018: Ultraviolet to Gamma Ray, ed. J.-W.~A. {den Herder}, S.~{Nikzad}, \& K.~{Nakazawa}, 1069929

\bibitem[{{Nanni} {et~al.}(2017){Nanni}, {Vignali}, {Gilli}, {Moretti}, \& {Brandt}}]{Nanni2017}
{Nanni}, R., {Vignali}, C., {Gilli}, R., {Moretti}, A., \& {Brandt}, W.~N. 2017, \aap, 603, A128

\bibitem[{{Rahman} {et~al.}(2024){Rahman}, {Sahayanathan}, \& {Subha}}]{Rahman2024}
{Rahman}, A.~A., {Sahayanathan}, S., \& {Subha}, P.~A. 2024, Journal of High Energy Astrophysics, 42, 156

\bibitem[{{Reynolds} {et~al.}(2024){Reynolds}, {Miller}, {Hodges-Kluck}, {Koss}, {Williams}, {Kara}, {Ptak}, {Mushotzky}, {Allen}, {Bauer}, {Bautz}, {Bodaghee}, {Burdge}, {Cappelluti}, {Cenko}, {Chartas}, {Chan}, {Corrales}, {Daylan}, {Falcone}, {Foord}, {Grant}, {Habouzit}, {Haggard}, {Herrmann}, {Kargaltsev}, {King}, {Kounkel}, {Lopez}, {Marchesi}, {McDonald}, {Meyer}, {Nynka}, {Okajima}, {Pacucci}, {Russell}, {Safi-Harb}, {Stassun}, {Falc{\~a}o}, {Walker}, {Wilms}, {Yukita}, \& {Zhang}}]{Reynolds2024}
{Reynolds}, C.~S., {Miller}, E.~D., {Hodges-Kluck}, E., {et~al.} 2024, in Society of Photo-Optical Instrumentation Engineers (SPIE) Conference Series, Vol. 13093, Space Telescopes and Instrumentation 2024: Ultraviolet to Gamma Ray, ed. J.-W.~A. {den Herder}, S.~{Nikzad}, \& K.~{Nakazawa}, 1309328

\bibitem[{{Sambruna} {et~al.}(2004){Sambruna}, {Gambill}, {Maraschi}, {Tavecchio}, {Cerutti}, {Cheung}, {Urry}, \& {Chartas}}]{Sambruna2004}
{Sambruna}, R.~M., {Gambill}, J.~K., {Maraschi}, L., {et~al.} 2004, \apj, 608, 698

\bibitem[{{Schwartz} {et~al.}(2000){Schwartz}, {Marshall}, {Lovell}, {Piner}, {Tingay}, {Birkinshaw}, {Chartas}, {Elvis}, {Feigelson}, {Ghosh}, {Harris}, {Hirabayashi}, {Hooper}, {Jauncey}, {Lanzetta}, {Mathur}, {Preston}, {Tucker}, {Virani}, {Wilkes}, \& {Worrall}}]{Schwartz2000}
{Schwartz}, D.~A., {Marshall}, H.~L., {Lovell}, J.~E.~J., {et~al.} 2000, \apjl, 540, 69

\bibitem[{{Schwartz} {et~al.}(2020){Schwartz}, {Siemiginowska}, {Snios}, {Worrall}, {Birkinshaw}, {Cheung}, {Marshall}, {Migliori}, {Wardle}, \& {Gobeille}}]{Schwartz2020}
{Schwartz}, D.~A., {Siemiginowska}, A., {Snios}, B., {et~al.} 2020, \apj, 904, 57

\bibitem[{{Shimwell} {et~al.}(2022){Shimwell}, {Hardcastle}, {Tasse}, {Best}, {R{\"o}ttgering}, {Williams}, {Botteon}, {Drabent}, {Mechev}, {Shulevski}, {van Weeren}, {Bester}, {Br{\"u}ggen}, {Brunetti}, {Callingham}, {Chy{\.z}y}, {Conway}, {Dijkema}, {Duncan}, {de Gasperin}, {Hale}, {Haverkorn}, {Hugo}, {Jackson}, {Mevius}, {Miley}, {Morabito}, {Morganti}, {Offringa}, {Oonk}, {Rafferty}, {Sabater}, {Smith}, {Schwarz}, {Smirnov}, {O'Sullivan}, {Vedantham}, {White}, {Albert}, {Alegre}, {Asabere}, {Bacon}, {Bonafede}, {Bonnassieux}, {Brienza}, {Bilicki}, {Bonato}, {Calistro Rivera}, {Cassano}, {Cochrane}, {Croston}, {Cuciti}, {Dallacasa}, {Danezi}, {Dettmar}, {Di Gennaro}, {Edler}, {En{\ss}lin}, {Emig}, {Franzen}, {Garc{\'\i}a-Vergara}, {Grange}, {G{\"u}rkan}, {Hajduk}, {Heald}, {Heesen}, {Hoang}, {Hoeft}, {Horellou}, {Iacobelli}, {Jamrozy}, {Jeli{\'c}}, {Kondapally}, {Kukreti}, {Kunert-Bajraszewska}, {Magliocchetti}, {Mahatma}, {Ma{\l}ek}, {Mandal}, {Massaro}, {Meyer-Zhao}, {Mingo}, {Mostert}, {Nair},
  {Nakoneczny}, {Nikiel-Wroczy{\'n}ski}, {Orr{\'u}}, {Pajdosz-{\'S}mierciak}, {Pasini}, {Prandoni}, {van Piggelen}, {Rajpurohit}, {Retana-Montenegro}, {Riseley}, {Rowlinson}, {Saxena}, {Schrijvers}, {Sweijen}, {Siewert}, {Timmerman}, {Vaccari}, {Vink}, {West}, {Wo{\l}owska}, {Zhang}, \& {Zheng}}]{Shimwell2022}
{Shimwell}, T.~W., {Hardcastle}, M.~J., {Tasse}, C., {et~al.} 2022, \aap, 659, A1

\bibitem[{{Shimwell} {et~al.}(2019){Shimwell}, {Tasse}, {Hardcastle}, {Mechev}, {Williams}, {Best}, {R{\"o}ttgering}, {Callingham}, {Dijkema}, {de Gasperin}, {Hoang}, {Hugo}, {Mirmont}, {Oonk}, {Prandoni}, {Rafferty}, {Sabater}, {Smirnov}, {van Weeren}, {White}, {Atemkeng}, {Bester}, {Bonnassieux}, {Br{\"u}ggen}, {Brunetti}, {Chy{\.z}y}, {Cochrane}, {Conway}, {Croston}, {Danezi}, {Duncan}, {Haverkorn}, {Heald}, {Iacobelli}, {Intema}, {Jackson}, {Jamrozy}, {Jarvis}, {Lakhoo}, {Mevius}, {Miley}, {Morabito}, {Morganti}, {Nisbet}, {Orr{\'u}}, {Perkins}, {Pizzo}, {Schrijvers}, {Smith}, {Vermeulen}, {Wise}, {Alegre}, {Bacon}, {van Bemmel}, {Beswick}, {Bonafede}, {Botteon}, {Bourke}, {Brienza}, {Calistro Rivera}, {Cassano}, {Clarke}, {Conselice}, {Dettmar}, {Drabent}, {Dumba}, {Emig}, {En{\ss}lin}, {Ferrari}, {Garrett}, {G{\'e}nova-Santos}, {Goyal}, {G{\"u}rkan}, {Hale}, {Harwood}, {Heesen}, {Hoeft}, {Horellou}, {Jackson}, {Kokotanekov}, {Kondapally}, {Kunert-Bajraszewska}, {Mahatma}, {Mahony}, {Mandal}, {McKean},
  {Merloni}, {Mingo}, {Miskolczi}, {Mooney}, {Nikiel-Wroczy{\'n}ski}, {O'Sullivan}, {Quinn}, {Reich}, {Roskowi{\'n}ski}, {Rowlinson}, {Savini}, {Saxena}, {Schwarz}, {Shulevski}, {Sridhar}, {Stacey}, {Urquhart}, {van der Wiel}, {Varenius}, {Webster}, \& {Wilber}}]{Shimwell2019}
{Shimwell}, T.~W., {Tasse}, C., {Hardcastle}, M.~J., {et~al.} 2019, \aap, 622, A1

\bibitem[{{Siemiginowska} {et~al.}(2003){Siemiginowska}, {Smith}, {Aldcroft}, {Schwartz}, {Paerels}, \& {Petric}}]{Siemiginowska2003}
{Siemiginowska}, A., {Smith}, R.~K., {Aldcroft}, T.~L., {et~al.} 2003, \apjl, 598, L15

\bibitem[{{Simionescu} {et~al.}(2016){Simionescu}, {Stawarz}, {Ichinohe}, {Cheung}, {Jamrozy}, {Siemiginowska}, {Hagino}, {Gandhi}, \& {Werner}}]{Simionescu2016}
{Simionescu}, A., {Stawarz}, {\L}., {Ichinohe}, Y., {et~al.} 2016, \apjl, 816, L15

\bibitem[{{Snios} {et~al.}(2021){Snios}, {Schwartz}, {Siemiginowska}, {Sobolewska}, {Birkinshaw}, {Cheung}, {Gobeille}, {Marshall}, {Migliori}, {Wardle}, \& {Worrall}}]{Snios2021}
{Snios}, B., {Schwartz}, D.~A., {Siemiginowska}, A., {et~al.} 2021, \apj, 914, 130

\bibitem[{{Spergel} {et~al.}(2007){Spergel}, {Bean}, {Dor{\'e}}, {Nolta}, {Bennett}, {Dunkley}, {Hinshaw}, {Jarosik}, {Komatsu}, {Page}, {Peiris}, {Verde}, {Halpern}, {Hill}, {Kogut}, {Limon}, {Meyer}, {Odegard}, {Tucker}, {Weiland}, {Wollack}, \& {Wright}}]{Spergel2007}
{Spergel}, D.~N., {Bean}, R., {Dor{\'e}}, O., {et~al.} 2007, \apjs, 170, 377

\bibitem[{{Spingola} {et~al.}(2020){Spingola}, {Dallacasa}, {Belladitta}, {Caccianiga}, {Giroletti}, {Moretti}, \& {Orienti}}]{Spingola2020}
{Spingola}, C., {Dallacasa}, D., {Belladitta}, S., {et~al.} 2020, \aap, 643, L12

\bibitem[{{Tavecchio}(2021)}]{Tavecchio2021}
{Tavecchio}, F. 2021, \mnras, 501, 6199

\bibitem[{{Tavecchio} {et~al.}(2000){Tavecchio}, {Maraschi}, {Sambruna}, \& {Urry}}]{Tavecchio2000}
{Tavecchio}, F., {Maraschi}, L., {Sambruna}, R.~M., \& {Urry}, C.~M. 2000, \apjl, 544, L23

\bibitem[{{Tavecchio} {et~al.}(2007){Tavecchio}, {Maraschi}, {Wolter}, {Cheung}, {Sambruna}, \& {Urry}}]{Tavecchio2007}
{Tavecchio}, F., {Maraschi}, L., {Wolter}, A., {et~al.} 2007, \apj, 662, 900

\bibitem[{{Tawa} {et~al.}(2008){Tawa}, {Hayashida}, {Nagai}, {Nakamoto}, {Tsunemi}, {Yamaguchi}, {Ishisaki}, {Miller}, {Mizuno}, {Dotani}, {Ozaki}, \& {Katayama}}]{Tawa2008}
{Tawa}, N., {Hayashida}, K., {Nagai}, M., {et~al.} 2008, \pasj, 60, S11

\bibitem[{{Vito} {et~al.}(2019){Vito}, {Brandt}, {Bauer}, {Calura}, {Gilli}, {Luo}, {Shemmer}, {Vignali}, {Zamorani}, {Brusa}, {Civano}, {Comastri}, \& {Nanni}}]{Vito2019}
{Vito}, F., {Brandt}, W.~N., {Bauer}, F.~E., {et~al.} 2019, \aap, 630, A118

\bibitem[{{Wachter} {et~al.}(1979){Wachter}, {Leach}, \& {Kellogg}}]{Wachter1979}
{Wachter}, K., {Leach}, R., \& {Kellogg}, E. 1979, \apj, 230, 274

\bibitem[{{Worrall}(2009)}]{Worrall2009}
{Worrall}, D.~M. 2009, The Astronomy and Astrophysics Review, 17, 1

\bibitem[{{Worrall} {et~al.}(2020){Worrall}, {Birkinshaw}, {Marshall}, {Schwartz}, {Siemiginowska}, \& {Wardle}}]{Worrall2020}
{Worrall}, D.~M., {Birkinshaw}, M., {Marshall}, H.~L., {et~al.} 2020, \mnras, 497, 988

\end{thebibliography}

\end{document}